\documentclass[
 reprint,
 amsmath,amssymb,
 aps,
floatfix,
]{revtex4-2}

\usepackage{graphicx}
\usepackage{dcolumn}
\usepackage{bm}
\usepackage[utf8]{inputenc}
\usepackage{pgf, tikz}
\usetikzlibrary{arrows, automata}
\usetikzlibrary{positioning}
\usepackage{amsmath}
\usepackage{amssymb}
\usepackage{amsthm}
\usepackage{comment}
\usepackage[linesnumbered,ruled,vlined]{algorithm2e}
\usepackage{xcolor}

\newcommand{\Npix}{N_{\mathrm{pix}}}
\newcommand{\lmax}{\ell_{\mathrm{max}}}
\newcommand{\R}{\mathbb{R}}
\newcommand{\E}{\mathbb{E}}
\newcommand{\eqdef}{:=}
\newcommand{\Ytilde}{\boldsymbol{\tilde{Y}}}
\newcommand{\Y}{\boldsymbol{Y}}
\newcommand{\NoiseMatrix}{\boldsymbol{N}}
\newcommand{\Beam}{\boldsymbol{B}}
\newcommand{\Gam}{\boldsymbol{\Gamma}}
\newcommand{\M}{\boldsymbol{M}}
\newcommand{\Sig}{\boldsymbol{\Sigma}}
\newcommand{\Q}{\boldsymbol{Q}}
\newcommand{\C}{\boldsymbol{C}}
\newcommand{\Var}{\mathrm{Var}}
\newcommand{\Cov}{\mathrm{Cov}}
\newcommand{\Corr}{\mathrm{Corr}}
\newcommand{\SNR}{\mathrm{SNR}}
\newcommand{\IAT}{\mathrm{IAT}}
\newcommand{\new}{\mathrm{new}}
\newcommand{\ESS}{\mathrm{ESS}}

\SetKwInput{KwInput}{Input}       
\SetKwInput{KwOutput}{Output}

\begin{document}

\preprint{APS/123-QED}

\title{Improved Gibbs samplers for Cosmic Microwave Background power spectrum estimation}

\author{Gabriel Ducrocq}
\email{gabriel.ducrocq@ensae.fr}
\affiliation{
    Institut Polytechnique de Paris, ENSAE Paris, CEDEX, 92247 Malakoff, France 
}

\author{Nicolas Chopin}
\email{nicolas.chopin@ensae.fr}
\affiliation{
    Institut Polytechnique de Paris, ENSAE Paris, CEDEX, 92247 Malakoff, France
}

\author{Josquin Errard}
\email{josquin@apc.in2p3.fr}
\affiliation{
   Université de Paris, CNRS, Astroparticule et Cosmologie, F-75013 Paris, France
}

\author{Radek Stompor}
\email{radek.stompor@in2p3.fr}
\affiliation{
CNRS-UCB International Research Laboratory, Centre Pierre Binétruy, IRL2007, CPB-IN2P3, Berkeley, US\\
Université de Paris, CNRS, Astroparticule et Cosmologie, F-75013 Paris, France
}

\begin{abstract}
{\color{black}
   We study different variants of the Gibbs sampler algorithm from the perspective of their applicability to the estimation of power spectra of the cosmic microwave background (CMB) anisotropies. These include approaches studied earlier in the CMB literature as well as new ones which are proposed in this work. We demonstrate all these variants on full and cut sky simulations and compare their performance, assessing both their computational and statistical efficiency. For this we employ a consistent comparison metric, an effective sample size (ESS) per second, commonly used in this context in the statistical literature. We show that one of the proposed approaches, referred to as Centered overrelax, which capitalizes on additional, auxiliary variables to minimize computational time needed per sample, and uses overrelaxation to decorrelate subsequent samples, performs better than the standard Gibbs sampler by a factor between one and two orders of magnitude in the nearly full-sky, satellite-like cases. It therefore potentially provides an 
interesting alternative to the currently favored approaches.}
\end{abstract}

\maketitle

\section{Introduction}
In the past few decades, the analysis of the Cosmic Microwave Background (CMB) has made a lot of progress. Numerous, novel and advanced statistical and numerical techniques have been proposed and implemented for virtually every step down the CMB data analysis pipeline. In particular, an entire slew of very diverse methods have been designed to produce estimates of the temperature or polarization power spectra or estimates of the cosmological parameters from a set of noisy CMB maps. \textcolor{black}{We can divide these in three broad categories. The first one includes the so-called pseudo-$C_{l}$ approaches, e.g.,~\cite{Upham2019, hamimeche_likelihood_2008, hamimeche_properties_2009, grain_polarized_2009, hivon_master_2002}, which compute the power spectra directly from the observed noisy  maps of the CMB sky. See~\cite{gerbino_likelihood_2020} for a review. The second category involves the maximum likelihood methods~\cite{Gjerloew2015, Tegmark2000}, which maximize the likelihood of the observed CMB maps with respect to the sought-after coefficients of the CMB power spectra. The third category comprises the Bayesian approaches using Markov chain Monte Carlo (MCMC) sampling methods, which directly target the posterior distribution of the estimated parameters, such as power spectra, given the observed data. A number of such techniques exist and some have been applied either for the power spectra or cosmological parameter estimation. These include the Metropolis-Hastings sampler, \cite{lewis_cosmological_2002, wraith_estimation_2009-2, eriksen_joint_2008}. the Hamiltonian Monte Carlo sampler \cite{taylor_fast_2008}, \cite{hajian_efficient_2007-3}, or the Gibbs sampler \cite{eriksen_power_2004, larson_estimation_2007, racine_cosmological_2016, jewell_markov_2009-2, wandelt_global_2004}. 
}

{\color{black}Out of those, the pseudo-$C_{l}$ methods are computationally very efficient but require careful characterisation of their statistical properties and a design of a corresponding pseudo-likelihood to allow for a meaningful interpretation of the estimated spectra. They are often a method of choice for the analysis of spectra at angular scales much smaller than the observed sky area, when such a pseudo-likelihood construction is more straightforward~\cite{gerbino_likelihood_2020}.

The maximum likelihood methods are statistically more robust. However, they are computationally heavy and typically require approximations to provide a meaningful description of the power spectrum likelihood. 
They are typically applicable only to downsized data sets providing constraints on the power spectra on large angular scales~\cite{Gjerloew2015, Tegmark2000}.

The MCMC sampling techniques can provide a robust description of the posterior distribution of the estimated spectra in the full range of angular scales. They do so by generating chains of samples which encode the statistical properties of the posterior. Some of these techniques can also cut significantly on the computational load of the maximum likelihood methods. Out of potential methods, Gibbs samplers has been found particularly well adapted to the context of the CMB power spectrum estimation, e.g.,~\cite{eriksen_power_2004, racine_cosmological_2016, jewell_markov_2009-2}, and this is a Gibbs sampler which is implemented in the most advanced, existing, Bayesian CMB power spectrum estimation code,~\cite{racine_cosmological_2016}. Gibbs samplers have also thorough statistical underpinning. In particular, the efficiency 
of the two-steps Gibbs sampler for linear hierarchical models, i.e., as used in the CMB context, have been extensively studied in the statistical literature, e.g.,~\cite{noauthor_fraction_nodate-2}.

The current implementations of the Gibbs sampler however remain computationally demanding. This often imposes practical limits on the number or the size of test and validation runs which can be afforded, and frequently requires approximations in modelling input CMB data in order to simplify the calculations. The speedup here comes however at the potential risk of increased statistical uncertainties, presence of biases or both, in the final results of such analyses. More efficient Gibbs algorithms are required in order to bypass such limitations.

There are two factors determining sampler's run time: the time needed to draw a single sample and the overall number of samples required to provide sufficient sampling of the posterior. How good the posterior sampling is, is best quantified by a number of effective, uncorrelated samples. This number is smaller, and typically much smaller, than the number of actual samples, which are usually correlated.
Stronger the correlations, the samples explore the volume of the posterior less efficiently, and consequently, more samples are needed to reach the same number of the effective samples. This effect is referred to in the statistical literature as a bad mixing of the algorithm~\cite{Robert2004}.

In this paper we present several new ideas aiming at enhancing the performance of the Gibbs sampler as applied to the CMB power spectrum estimation. These include methods, which aim at cutting the number of actual samples required to characterize reliably the posterior, i.e., improving mixing properties of the algorithms, as well as methods which attempt to trade the time needed to compute samples for their number, potentially leading to a net gain in the overall performance. The third possibility of improving numerical algorithms and their implementation to cut on the computational time of each sample is not considered in this work. To compare the different methods, we evaluate the number of effective, uncorrelated samples which these methods can produce per unit time. This metric, referred to as an effective sample size per second, ESS, is defined and discussed in Sect.~\ref{Experiments}.

We organize this paper as follows. In Section~\ref{DataModel} we review the adopted data model and introduce the basic formalism. In Section~\ref{CenteredGibbs} we present the standard Gibbs sampler as considered in early CMB power spectrum estimation literature and discuss its deficiencies. In Section~\ref{NonCenteredGibbs} and~\ref{interweaving} we discuss techniques aiming at decreasing the number of necessary samples, while keeping the sample computations unchanged. In Section~\ref{CRStep}, we discuss ways to suppress time needed for the single sample computations, compensating by an increased number of samples. Finally, in Section~\ref{Experiments}, we describe our experiments and compare the performance of all the presented Gibbs variants. We show that on nearly full-sky, satellite-like data, one of the proposed algorithms performs (in terms of the effective sample size per second) one order of magnitude better on the $EE$ power spectrum and two orders of magnitude better on the $BB$ power spectrum than our baseline algorithm. We conclude our findings in Section~\ref{Conclusion}.
}

\section{Basic formalism}\label{DataModel}
\subsection{Data model}
{\color{black}
We assume throughout this work that the input data set consists of noisy maps including only the CMB signal and we focus on the estimation of its power spectra from such maps. The maps typically cover only part of the entire sky and can be of one, two or three Stokes parameters, corresponding to the total intensity, $I$, or Stokes parameters, $Q$ and $U$ only, or all three Stokes parameters, $I$, $Q$, and $U$, respectively. The CMB signal is assumed to be Gaussian, with the covariance given by matrix $\C$. The noise in the maps is also Gaussian with the covariance given by $\NoiseMatrix$.}
The data model underlying the maps is therefore hierarchical, (see~\cite{eriksen_joint_2008, eriksen_power_2004, wandelt_global_2004, chu_cosmological_2005}), and reads,
\begin{equation}\label{CenteredModel}
\begin{array}{ccc}
    \{\mathbf{C}_{\ell}\} & \sim & p_{0},\cr
    s|\{\mathbf{C}_{\ell}\} & \sim & \mathcal{N}(0, \C),\cr
    d|s & \sim & \mathcal{N}(\Ytilde s, \NoiseMatrix), \cr
\end{array}
\end{equation}
Here,
$\sim$ denotes a sample drawn from the distribution on the right hand side. $p_{0}$ is the flat prior and $\mathcal{N}(m, \boldsymbol{\Sigma})$ denotes the Gaussian distribution with mean $m$ and covariance  $\boldsymbol{\Sigma}$.

$\{\mathbf{C}_{\ell}\}_{2\leq \ell \leq \lmax}$ stands here for a set of all relevant power spectra coefficients numbered by a multiple number, $\ell$, (with the monopole and dipole ignored in the case of total intensity power spectrum). 
These uniquely define the CMB covariance matrix, $\C$. Each $\mathbf{C}_\ell$ can be a number, i.e., in the case of the total intensity maps, or a matrix, in the case of multiple Stokes parameter maps, as elaborated below, equation~\eqref{eqn:powSpecMat}.

$s$ 
is the sky map expressed in the spherical harmonic basis. Hereafter, we follow the convention that we use two real numbers (one for the coefficients with $m=0$) corresponding to the real and imaginary part of the sky harmonic coefficients, instead of a single complex one, and separate them into two real vectors, see \cite{Seljebotn2010} for an extensive justification. 

$\Ytilde$ is the product of the spherical harmonics synthesis matrix $\Y$ and the (diagonal) Gaussian beam matrix $\Beam$, in the spherical harmonic domain, which is assumed diagonal corresponding to an axially symmetric beam. Consequently, $\Ytilde\, s$ stands for the beam-smoothed CMB map in the pixel domain computed from the harmonic coefficients, $s$, drawn from the Gaussian distribution with covariance $\C$. {\color{black}The map, $\Ytilde\, s$, corresponds only to the observed part of the sky for which the data, as defined by the data vector, $d$, are available and in general it will cover only limited area of the full sky.}

We assume that for a full sky map,
\begin{equation}
\dfrac{4\pi}{\Npix}\Y^{T}\Y = \boldsymbol{I},
\end{equation}
where $\boldsymbol{I}$ denotes the identity matrix. This means that the adapted pixelization used to descretize the map objects is such that all the spherical harmonics all the way up the band limit, $\lmax$, are orthogonal on the grid made of the pixel centers.

The data vector, $d \in \R^{\Npix}$, is the noisy sky map in the pixel domain and $\Npix$ is the number of pixels of the map.

We note that the noise covariance, $\NoiseMatrix$, is given in the pixel domain and assumed hereafter to be diagonal (though not necessarily to be proportional to the unit matrix). In contrast, the signal covariance, $\C$, is defined in the harmonic domain, and is block diagonal (diagonal in the case of total intensity only). For example, in the case of inference on total intensity and polarization, the blocks of the signal covariance are,
\begin{equation}
\boldsymbol{C}_{\ell} = 
\begin{pmatrix} 
C_{\ell}^{TT} & C_{\ell}^{TE} & C_{\ell}^{TB} \\
C_{\ell}^{TE} & C_{\ell}^{EE} & C_{\ell}^{EB}\\
C_{\ell}^{TB} & C_{\ell}^{EB} & C_{\ell}^{BB} \\
\end{pmatrix}
\label{eqn:powSpecMat}
\end{equation}
where $\{C_{\ell}^{TT}\}$, $\{C_{\ell}^{EE}\}$,$\{C_{\ell}^{BB}\}$, $\{C_{\ell}^{TE}\}$, $\{C_{\ell}^{TB}\}$, $\{C_{\ell}^{EB}\}$ are the temperature, E-mode, B-mode and cross correlations power spectra. For the standard, parity-invariant cosmology, adopted in this work, $\{C_{\ell}^{TB} \} = \{C_{\ell}^{EB}\} = 0$.
In the rest of this paper, for simplicity, we will drop the dependency of the signal covariance matrix on the power spectrum and define $\C \eqdef \C(\{\mathbf{C}_{\ell}\})$.

For the sake of transparency, hereafter we present our algorithms specialized for the case of the total intensity as the generalization to include polarization is straightforward, see~\cite{larson_estimation_2007} for example.
We however include polarization in all our numerical experiments in Section~\ref{Experiments}.

\subsection{Likelihood}
If the observed data are normally distributed given the power spectrum, the likelihood of the observed data reads,
\begin{equation}\label{likelihood}
\mathcal{L}(d|\{C_{l}\}) \propto \dfrac{\exp\left\{-(1/2)d^{T} (\Ytilde \mathbf{C} \Ytilde^{T} + \NoiseMatrix)^{-1}d\right\}} {|\Ytilde \mathbf{C}\Ytilde^{T} + \NoiseMatrix|^{1/2}},
\end{equation}
where $|\dots|$ denotes the absolute value of the determinant of a matrix, and $\Ytilde \mathbf{C} \Ytilde^{T}$ is the signal covariance of the observed, therefore either full or cut-sky, map in the pixel domain.

The full covariance matrix of this likelihood is dense in the pixel domain and, in the case of partial sky coverage and noise covariance matrix not proportional to the identity, is also dense in the harmonic domain, therefore inverting it and computing its determinant is time consuming as soon as the dimension, i.e., the number of the observed sky pixels, is high. Hence the computation of the likelihood, and therefore the maximum likelihood approach, becomes quickly prohibitive. We can however rely on the Bayesian approach instead.

\subsection{Bayesian approach}
Adopting the Bayesian viewpoint and putting an improper flat prior $p_{0}(\{C_{l}\})$ on the power spectrum, we can derive the posterior distribution of the power spectrum coefficients,
\begin{equation}\label{posterior}
       \pi(\{C_{l}\}|d) \propto \mathcal{L}(d|\{C_{l}\}).
\end{equation}
Unfortunately, evaluating this posterior is as computationally involved as the computation of the likelihood and making application of the sampling algorithms difficult or, as in the case of Metropolis-Hastings sampler, directly infeasible.

To bypass this difficulty we can augment our data model and consider a joint posterior over the power spectrum and sky map, as done in previous works \cite{eriksen_joint_2008, eriksen_power_2004, wandelt_global_2004, chu_cosmological_2005, larson_estimation_2007}:
\begin{equation}\label{eqn:augmentedBayes}
\pi(\{C_{l}\}, s | d) \propto p(d|s) p(s|\{C_{l}\}).
\end{equation}
where 
\begin{equation}
\log p(d|s) = -(d - \Ytilde s)^{T}\NoiseMatrix^{-1}(d-\Ytilde s)/2 + c_{1}
\end{equation}
and 
\begin{equation}
\log p(s|\{C_{l}\}) = - s^{T}\C^{-1}s/2- \log |\C|/2 + c_{2}
\end{equation}
with $c_{1}, c_{2}$ -- real valued constants. We note that $s$ denotes the set of spherical harmonic coefficients and is therefore equivalent to the full sky map in the pixel domain, notwithstanding the fact that available data, $d$, may correspond only to a partial sky. Consequently, the number of elements of $s$ can be much larger than the number of the data points collected in $d$.
The elements of $s$ are referred to as latent variables, as they are introduced to facilitate the computation and will be eventually discarded.
As their covariance matrix, $\C$, is very structured, its determinant and its inverse are both straightforwardly computable.
Hence, we could apply the Metropolis-Hastings algorithm to this joint distribution, however, we do not expect it to be efficient due to the high dimensionality of the problem and the strong correlations between the variables. However, as first proposed in~\cite{jewell_application_2004, wandelt_global_2004}, we can sample from the respective conditional posterior distributions of this joint posterior and can apply a Gibbs sampler instead. We discuss this in detail in the next section.

We note that in general using an improper prior distribution may lead to an improper posterior distribution - that is one with infinite mass - creating problems for MCMC algorithms as discussed in the statistical literature, e.g.,~\cite{hobert_effect_1996}. However, in our application and in the case of full-sky data it can be shown, see appendix~\ref{PriorsAppendix}, that the improper flat prior, $p(\{C_{l}\})$, results in a proper posterior distribution. This is consistent with the previous CMB literature on MCMC applications,~\cite{eriksen_joint_2008, eriksen_power_2004, wandelt_global_2004, chu_cosmological_2005, larson_estimation_2007}, which have reported no pathological cases.
In contrast, as also shown in appendix~\ref{PriorsAppendix}, using Jeffrey's prior~\cite{Harold1946} on this model, as also suggested in some previous CMB works, e.g \cite{larson_estimation_2007} and \cite{eriksen_joint_2008}, leads to an improper posterior distribution in the case of full sky observation and thus results in a non-valid MCMC algorithm.
Given that, and following the accepted convention in the field, we adopt the improper flat prior on the power spectrum throughout this work.

\section{Gibbs Sampling}\label{CenteredGibbs}
\subsection{The algorithm}
The principle of Gibbs sampling for data augmentation is to sample iteratively from the conditional distributions of the parameters and the latent variables,
see, e.g.,~\cite{Tanner1987}. Algorithm~\ref{AlgoCentered} shows one iteration of this technique applied to the joint posterior distribution in equation~\eqref{eqn:augmentedBayes}.

\begin{algorithm}\label{AlgoCentered}
\caption{Iteration $t$ of Gibbs sampling for Data Augmentation}
    \KwInput{$(\{C_{\ell}\}_{t}, s_{t})$}
    \KwOutput{$(\{C_{\ell}\}_{t+1}, s_{t+1})$}
    $s_{t+1} \sim p(s|d, \{C_{\ell}\}_{t})$ \tcp{Constrained Realization step}
    
    $\{C_{\ell}\}_{t+1} \sim p(\{C_{\ell}\}|d, s_{t+1})$ \tcp{Power Sampling step}
\end{algorithm}

The first step of drawing a sample of the sky signal, $s_{t+1}$, given the data and the power spectrum is called the constrained realization step. The second step is the power spectrum sampling step as it draws a sample of the power spectrum given the data and the sky signal. This type of algorithms has been widely used for CMB data analysis \cite{eriksen_joint_2008, eriksen_power_2004, wandelt_global_2004,chu_cosmological_2005, larson_estimation_2007}. The hierarchical data model underlying Algorithm~\ref{AlgoCentered} can be represented graphically by a directed acyclic graph (DAG) shown in Figure~\ref{fig:DAGCentered}.

\subsection{Constrained Realization step}\label{CRSection}
The distribution of the sky map, conditional on the observed map and the power spectrum, is given by,
\begin{equation}
\label{eq:constrainredRealConditional}
s|d, \{C_{\ell}\} \sim \mathcal{N}(\mu, \Sig)
\end{equation}
where 
\begin{eqnarray}
\Sig && \eqdef \Q^{-1}= (\Ytilde^{T}\NoiseMatrix^{-1}\Ytilde + \C^{-1})^{-1}\nonumber \\ \mu && \eqdef \Sig \Ytilde^{T}\NoiseMatrix^{-1}d.
\label{eqn:constRealPars}
\end{eqnarray}
However, in the case of an inhomogeneous noise and/or an incomplete sky coverage, the covariance matrix in equation~\eqref{eq:constrainredRealConditional}, $\Sig$, is dense and highly dimensional. Hence it is costly to invert it or to compute its Cholesky decomposition.

In order to sample from this Gaussian distribution, we can rely instead on an algorithm proposed in the CMB context in~\cite{wandelt_global_2004} and known in the statistical literature as the Perturbation-Optimization algorithm~\cite{Orieux2012}. The steps are:
\begin{itemize}
    \item Draw $w_{0}, w_{1} \sim \mathcal{N}(0, \boldsymbol{I})$
    \item Solve for $x$: 
  \ \   \begin{multline}\label{eq:PCG}
(\Ytilde^{T}\NoiseMatrix^{-1}\Ytilde + \C^{-1})x = \Ytilde^{T}\NoiseMatrix^{-1/2} w_{0} + \C^{-1/2} w_{1}\\ + \Ytilde^{T}\NoiseMatrix^{-1}d \ \ \ \ \ 
\end{multline}
\end{itemize}
where $\boldsymbol{M}^{1/2}$ denotes any matrix satisfying: 
\[\boldsymbol{M} = \boldsymbol{M}^{1/2}(\boldsymbol{M}^{1/2})^{T}.
\]
Obviously, the right-hand term of equation~\eqref{eq:PCG} is a normal variable with distribution $\mathcal{N}(\Ytilde^{T}\NoiseMatrix^{-1}d, \mathbf{Q})$ and the solution of this system is a random variable drawn from the distribution in equation~\eqref{eq:constrainredRealConditional}.
Since this system may be very high-dimensional and badly conditioned, in practice the system in equation~\eqref{eq:PCG} is solved using an iterative solver such as preconditioned conjugate gradient (PCG) algorithm.
This indeed has been the standard way of making the constrained realization step in the context of CMB data analysis, see \cite{eriksen_joint_2008, eriksen_power_2004, wandelt_global_2004, chu_cosmological_2005, larson_estimation_2007, jewell_markov_2009-2}, however, see, e.g.,~\cite{elsner_wandelt_2013}, for alternative solvers, and~\cite{Papez2018} for their comparison. In the following, we introduce the Truncated Perturbation-Optimization (TPO) algorithm, which is a Perturbation-Optimization algorithm using an iterative method to solve the linear system, which is terminated after a predetermined number of iterations or reaching a precision threshold and therefore potentially failing to attain sufficient accuracy.

\subsection{Power spectrum sampling}
\begin{figure}[htpb]
\includegraphics[scale = 0.6]{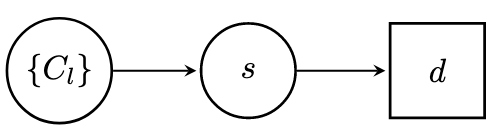}
    \caption{\textcolor{black}{Directed acyclic graph of model~\ref{CenteredModel}. Circles and squares represent unobserved and observed variables respectively. Plain arrows represent stochastic dependence.}}\label{fig:DAGCentered}
\end{figure}

The second step of the Gibbs sampler in Algorithm~\ref{AlgoCentered} consists in sampling the power spectrum conditionally on the sky signal, $s$, and the observed data, $d$. As visualized in Figure~\ref{fig:DAGCentered}, the sampling is in fact independent on the data as $p(\{C_\ell\} | s, d) = p(\{C_\ell\} | s)$ and given by,
    \begin{eqnarray}
    p(\{C_{l}\}|s) &\propto & \dfrac{\exp\left\{-(1/2) s^{T}\C^{-1}s\right\}}{|\C|^{1/2}}.
\end{eqnarray}
In the case of either temperature or polarization, separately, this corresponds to a product of inverse gamma distributions~\cite{eriksen_joint_2008, eriksen_power_2004, wandelt_global_2004, chu_cosmological_2005, larson_estimation_2007},
    \begin{eqnarray}
    p(\{C_{\ell}\}|s) &\propto & \prod_{\ell=2}^{\lmax} \dfrac{\exp\left\{-(2\ell+1)\sigma_{\ell}/2C_{\ell}\right\}}{C_{\ell}^{(2\ell+1)/2}},
\end{eqnarray}
where,
\[\sigma_{\ell} \eqdef \dfrac{1}{2\ell+1}\sum_{-\lmax\leq m \leq \lmax} |a_{\ell,m}|^{2}\] is the empirical power spectrum.
In the case of temperature and polarization, we must instead sample from independent inverse Wishart distributions, see~\cite{eriksen_joint_2008, eriksen_power_2004, wandelt_global_2004, chu_cosmological_2005, larson_estimation_2007}. 
Hereafter, we continue presenting the formalism for the total intensity case and include polarization only in the numerical experiments in Section~\ref{Experiments}.

\subsection{Shortcomings}
While being straightforward to implement and tuning-free, this algorithm has two major issues that can prohibit its application in many cases of interest. These are: the high computational cost of the constrained realization step and the strong correlations between the sky map and the power spectrum. This is this last property, referred to in the statistical literature as bad mixing of the algorithm, which drives the number of samples needed to sample the full volume of the posterior. Both these factors tend to inflate the overall computational time of the algorithm potentially limiting its applicability. We discuss each of them in more detail below.

\subsubsection{Constrained realizations}\label{CR}
The resolution of the system in equation~\eqref{eq:PCG} is costly in general. Depending on the preconditioner that is being used, between $\mathcal{O}(350) $ and $\mathcal{O}(1000)$ spherical harmonics transforms were required for a WMAP-like experiment, with eight frequency bands,
see \cite{eriksen_power_2004}. In this work, we find that the resolution of the system takes $\mathcal{O}(240)$ spherical harmonics transforms for a lower resolution, LiteBIRD-like experiment with an 80\% Planck galactic mask, assuming the standard, Block-Jacobi preconditioner. 

More sophisticated preconditioners could speed up the convergence, however they typically require expensive precomputation and extra time to apply them. These can significantly offset any gain in the number of iterations they may bring. We note that in principle we need highly accurate solutions, what exacerbates the computational problem. The high accuracy is necessary to ensure that the solutions are really drawn from the required distribution. So while it may be tempting to compromise on the solution precision in the interest of the time, for low accuracy solutions, we may not even know what is the true underlying distribution they have been effectively drawn from, potentially invalidating the entire procedure. 

This is a real issue for the Gibbs sampler, since if we are not sampling from the correct conditional distributions at each iteration, we have no idea what effective joint distribution the Gibbs sampler is simulating from or even whether this distribution exists at all.

\subsubsection{Power spectrum sampling}
The second problem concerns the sampling of the power spectrum conditional on the sky map, that is, the second step of our Gibbs sampler.

We define the lag-1 autocorrelation \textcolor{black}{for any function $f$ with finite second order moment under $\pi$, i.e., for which $\int f^2(x)\, \pi(x)\, dx$ is finite, as}
\begin{equation}
    \gamma_{f} = \dfrac{\Cov(f(\{C_{\ell}\}_{0}), f(\{C_{\ell}\}_{1})|d)}{\Var(f(\{C_{\ell}\})|d)},
\end{equation}
where $\{C_{\ell}\}_{0} \sim \pi(\{C_{\ell}|d)$ and $\{C_{\ell}\}_{1}$ are two consecutive power spectrum samples computed at stationarity.
It has been shown in the statistical literature~\cite{noauthor_fraction_nodate-2} that in the case of data augmentation as in the case under consideration, at stationarity, the lag-1 autocorrelation, $\gamma_{f}$, can be expressed as,
\begin{equation}\label{IIR}
    \gamma_{f} = 1 - \dfrac{\boldsymbol{\E}\{\Var(f(\{C_{l}\})|s, d)|d\}}{\Var(f(\{C_{l}\})|d)}.
\end{equation}
\textcolor{black}{Following the statistical literature results, see \cite{liu_covariance_1995} and \cite{J.S.1994}}, it can be shown that the geometric rate of convergence of the Gibbs sampler -- see equation~\eqref{def:geometicRate} in Appendix~\ref{mixing} for a definition, $\gamma$, reads,
\begin{equation}\label{Corr}
\gamma = \sup_{f} \gamma_{f} = \{\sup_{f,g} \Corr(f({C_{l}}), g(s)|d)\}^{2}
\end{equation}
where the supremum is taken \textcolor{black}{over all functions with finite variance under $\pi$}, and $\Cov$, $\Var$, $\Corr$, and $\boldsymbol{\E}$ stand respectively for covariance, variance, correlation, and expectation value of the arguments. Equation~(\ref{IIR}) 
{\color{black}
shows that the lag-1 autocorrelation is determined by the fraction of the ``conditional variance'' over the posterior variance.}
If the conditional variance of the power spectrum given the sky is very small compared to the posterior variance of the power spectrum, then the lag-1 autocorrelation is high, leading to an inefficient sampling of the posterior and the bad mixing of the algorithm.
equation~\eqref{Corr} states that this happens when $\{C_{\ell}\}$ and $s$ are highly correlated.

This is actually intuitive: when the variance of the conditional distribution is small compared to the posterior one, sampling from this conditional distribution will make only ``small steps'', changing very little the power spectrum compared to the full range of potential posterior values. This in turn will lead to a small change as compared to the full posterior when sampling the signal conditionally on the power spectrum and so on. 
Consequently, the algorithm will not explore the posterior distribution efficiently.

\begin{figure}[htpb]
\begin{center}
\advance\leftskip-0.5cm
\includegraphics[scale=0.27]{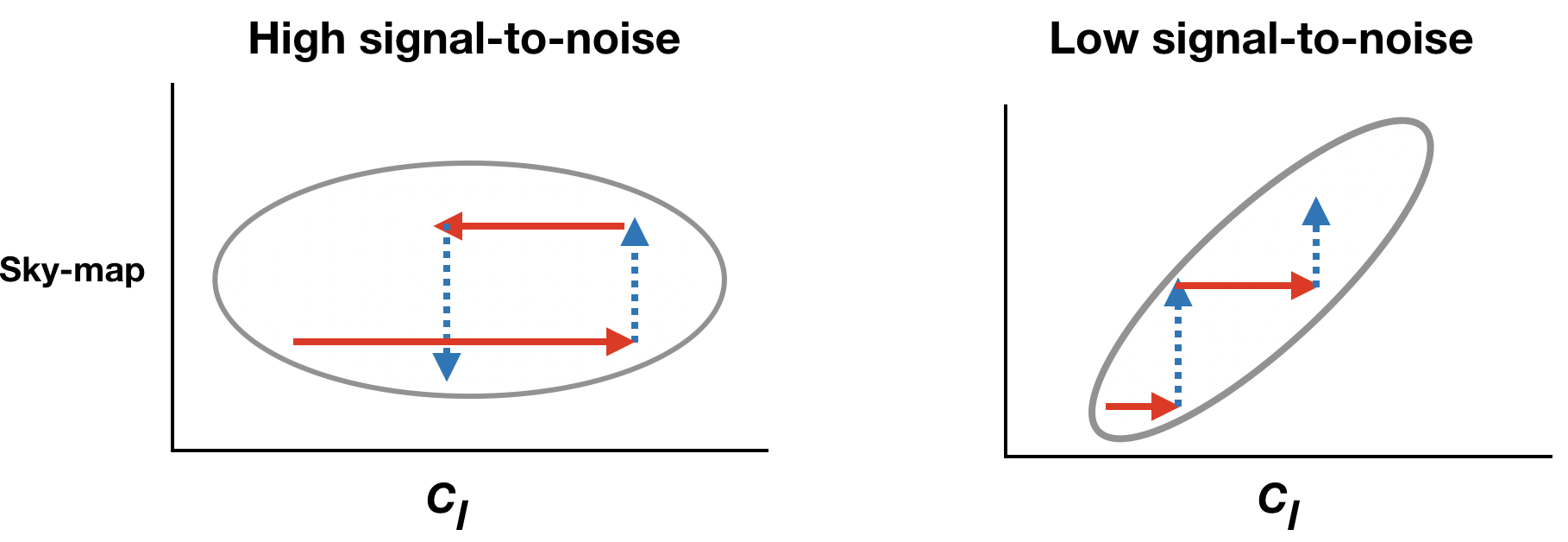}
\caption{{\color{black}Example of a sequence of consecutive samples of the Gibbs sampler in the centered parametrization. For low signal-to-noise power spectrum coefficients, shown in the right panel, the sky map and the power spectrum are strongly correlated. This leads to {\em the bad mixing of the algorithm} in this regime and a large number of samples is needed to explore the posterior marginals for such coefficients. This is not the case of the high signal-to-noise coefficients as shown in the left panel. Here, the correlations are small and {\em the resulting mixing of the algorithm is good} with many fewer samples needed to explore the posterior.
In both panels the red plain arrows depict sampling of the power spectrum given the sky map and the blue dotted arrows sampling the sky map given the power spectrum.} 
}\label{fig:CenteredMoves}
\end{center}
\end{figure}

Unfortunately, we encounter this problem in our application.
{\color{black}
Indeed, let us consider the case where we observe the full sky and have an isotropic noise covariance matrix: in this case the matrix $(\mathbf{C} + \Ytilde^{T}\mathbf{N}\Ytilde)^{-1}$ is diagonal in the harmonic domain and the posterior distribution is a product of inverse translated Gamma distribution and we have roughly:
\begin{align*} 
        \Var(C_{\ell}|d) &\propto (C_{\ell} + N_{\ell})^{2}  \\ 
        \Var(C_{\ell}|s) &\propto C_{\ell}^{2},
\end{align*}
where the noise power spectrum, $N_\ell$, includes the beam effects.
Hence, the lag-1 autocorrelation for multipole $\ell$ reads,
\begin{eqnarray}
\gamma_f^{\left(\ell\right)} \approx 1 - \left(\frac{C_{\ell}}{C_{\ell} + N_{\ell}}\right)^2 = 1 - \left(\frac{\SNR_{\ell}}{\SNR_{\ell} + 1}\right)^2,
\end{eqnarray}
where $\SNR_\ell$ stands for the signal-to-noise ratio of the power spectrum coefficient corresponding to multipole $\ell$, defined as,
\begin{equation}
\SNR_{\ell} = \dfrac{C_{\ell}}{N_{\ell}}.
\label{eq:snr}
\end{equation}
This shows that the lag-1 autocorrelations are determined by the SNRs of respective power spectrum coefficients. Consequently, they are going to be different for different coefficients. In the CMB power spectrum estimation we estimate a broad range of $C_\ell$, as defined on one end by the size of the observed sky patch and by the instrument resolution on the other, and which will therefore typically span a large range of SNR values, from high signal-to-noise cases with $\SNR \gg 1$ to the low signal-to-noise ones with $\SNR < 1$ and passing through those for which $\SNR \sim 1$.

We can now see that for low signal-to-noise coefficients the standard Gibbs sampler will not sample their respective marginal distributions efficiently. This is because for $\SNR_\ell \ll 1$, $\gamma_f^{\left(\ell\right)} \sim 1$, what indicates, following on the previous discussion, that the posterior variance of the respective marginal will be much bigger than its conditional variance.  From equation~\eqref{Corr} this is related to the fact that the correlation between the power spectrum coefficients and the sky maps in this regime are strong.

For high signal-to-noise cases, $\gamma_f^{\left(\ell\right)} \sim 0$, and the conditional and posterior variances are comparable, the correlations between the power spectra and the sky are expected to be significantly lower, and we expect that the algorithm will perform well for these coefficients, or in the statistical jargon, that it will mix well for these marginal distributions. 

All these observations are graphically summarized in Fig.~\ref{fig:CenteredMoves}.
}

\section{Non Centered Gibbs sampling}\label{NonCenteredGibbs}
\subsection{Algorithm}
\begin{figure}[htpb]
\includegraphics[scale = 0.6]{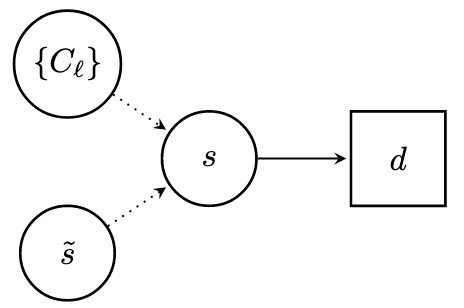}
\caption{\textcolor{black}{Directed acyclic graph of the model in equation~\eqref{NonCenteredModel}. Circles and squares represent unobserved and observed variables respectively. Plain arrows represent stochastic dependence. Dashed arrows represent deterministic dependence.}}\label{fig:DAGNonCentered}
\end{figure}

To circumvent this problem, we reparametrize the model in equation~\eqref{CenteredModel} to break the dependencies between the signal and the power spectrum. Such an approach was studied in the statistical literature, e.g.,~\cite{noauthor_non-centered_nodate-2, papaspiliopoulos_general_2007, agapiou_analysis_2014} and the CMB context in~\cite{jewell_markov_2009-2}. The new model reads,
\begin{equation}
\begin{array}{ccc}
    \{C_{l}\} &\sim& p_{0}\\
    \tilde{s} &\sim& \mathcal{N}(0, \boldsymbol{I})\\
    d &=& \Ytilde \C^{1/2}\tilde{s} + n
\end{array}\label{NonCenteredModel}
\end{equation}
where $n\sim\mathcal{N}(0, \NoiseMatrix)$ and $\boldsymbol{I}$ is the identity matrix of dimension $(\lmax+1)^{2} - 4$. We plot its directed acyclic graph representation in Figure~\ref{fig:DAGNonCentered}.

In this parametrization, the power spectrum, $\{C_\ell\}$, and the signal, $\tilde{s}$, are now independent a priori and all the posterior correlations come from the likelihood of the model.

In order to sample from that model we are also using a Gibbs sampler. Algorithm~\ref{NonCenteredAlgo} shows one iteration of the algorithm.

\begin{algorithm}\label{NonCenteredAlgo}
\caption{Iteration $t$ of the non centered Gibbs sampler}
    \KwInput{$(\{C_{\ell}\}_{t}, \tilde{s}_{t})$}
    \KwOutput{$(\{C_{\ell}\}_{t+1}, \tilde{s}_{t+1})$}
    $\tilde{s}_{t+1} \sim p(\tilde{s}|d, \{C_{\ell}\}_{t})$\\
    $\{C_{\ell}\}_{t+1} \sim p(\{C_{\ell}\}|d, \tilde{s}_{t+1})$ \\
\end{algorithm}

The first step is implemented like the first step of the centered Gibbs sampler except that at its conclusion we change the parametrization: we simulate $s_{t+1} \sim p(s|d, \{C_{\ell}\}_{t})$ and then set $\tilde{s}_{t+1} = \C_{t}^{-1/2}s_{t+1}$.
The second step, however, is different. This is because the power spectrum and the observed sky map are not independent when conditioned on the signal map. The second conditional density takes the following form,
\begin{multline}\label{NCPS}
\log p(\{C_{l}\}|\tilde{s}, d) = -\dfrac{1}{2}(d-\Ytilde \C^{1/2}\tilde{s})^{T} \NoiseMatrix^{-1}(d-\Ytilde \C^{1/2}\tilde{s})\\ + c
\end{multline}
where $c$ is a constant.
Since we are unable to sample directly from this conditional, we rely on a Metropolis step. This is implemented as follows, 
\begin{center}
\begin{itemize}
    \item Propose $\{C_{\ell}\}_{\new}\sim q(\cdot |\{C_{\ell}\}_{t})$
    \item Set $\{C_{\ell}\}_{t+1} = \{C_{\ell}\}_{\new}$ with probability \[ r = \min(1, \alpha)\], where
    \begin{multline}
        \alpha =  \dfrac{\exp\left\{-(d-\Ytilde \C_{\new}^{1/2}\tilde{s}_{t})^{T} \NoiseMatrix^{-1}(d-\Ytilde \C_{\new}^{1/2}\tilde{s}_{t})/2\right\}}{\exp\left\{-(d-\Ytilde \C_{t}^{1/2}\tilde{s}_{t})^{T} \NoiseMatrix^{-1}(d-\Ytilde \C_{t}^{1/2}\tilde{s}_{t})/2\right\}}\\\times
        \dfrac{q(\{C_{\ell}\}_{t}|\{C_{\ell}\}_{\new})}{q(\{C_{\ell}\}_{\new}|\{C_{\ell}\}_{t})},
    \end{multline}
    otherwise set $\{C_{\ell}\}_{t+1} = \{C_{\ell}\}_{t}$.
\end{itemize}
\end{center}
Here $q(.|\{C_{\ell}\}_{t})$ is the proposal distribution assumed to be normal with a diagonal covariance matrix, centered in $\{C_{\ell}\}_{t}$, whose marginals are truncated to
real positive numbers.
This algorithm has already been implemented in the context of CMB data analysis in~\cite{jewell_markov_2009-2}.
In addition, since the problem is very high-dimensional, we decompose $\{C_{\ell}\}$ into disjoint subsets and we sample each them in turn, one-by-one, while keeping all others fixed following the approach of~\cite{jewell_markov_2009-2}. Consequently, we are implementing a Gibbs sampler targeting distribution in equation~\eqref{NCPS}, however each Gibbs step is performed thanks to the Metropolis step. We also follow~\cite{jewell_markov_2009-2} in order to tune the diagonal elements of the covariance matrix of the proposal distribution, $q$.

\subsection{Shortcomings}
We can already expect this algorithm to suffer from two main shortcomings.
First, we still have to solve a high-dimensional linear system, as described in Section~\ref{CR}. The problems are the same, namely, the high computational cost of the algorithm and the fact that it may not always converge to a solution which is sufficiently accurate.

\begin{figure}[htpb]
\begin{center}
\advance\leftskip-0.5cm
\includegraphics[scale=0.27]{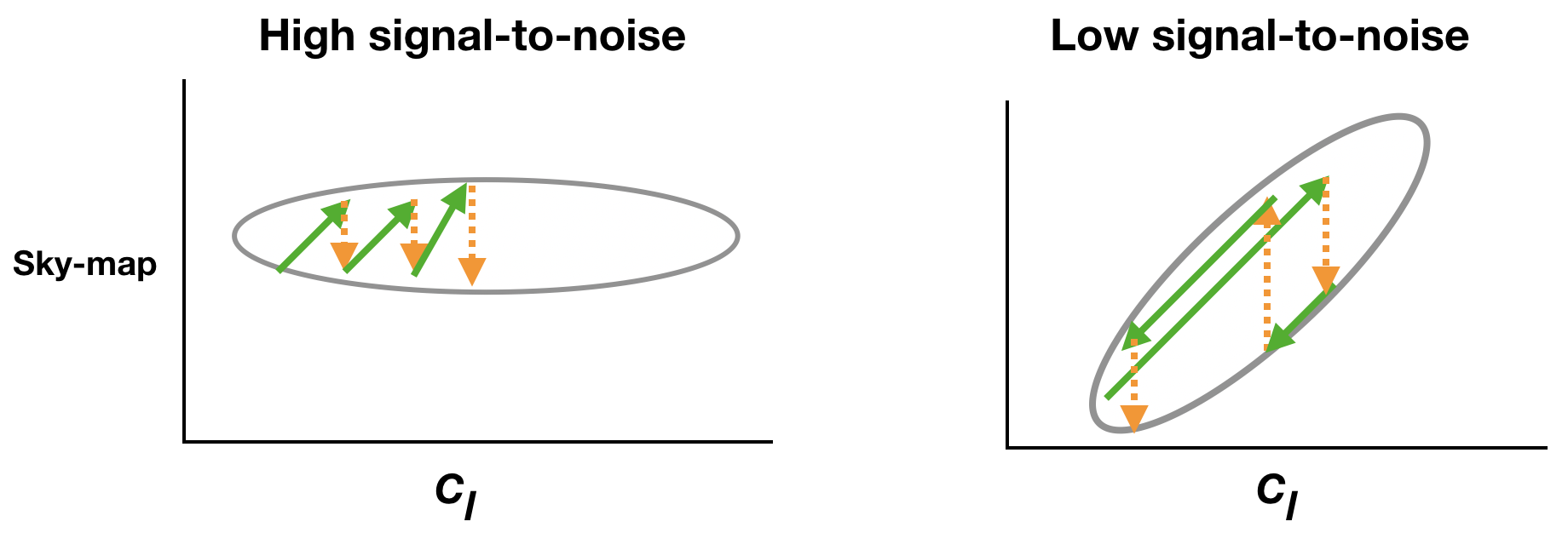}
\caption{Example of a sequence of samples of the non-centered Gibbs sampler. For low signal-to-noise power spectrum coefficients, the sky map and the power spectrum are strong correlated, right panel. This is not the case of the high signal-to-noise coefficients, left panel. The green plain arrows depict sampling the power spectrum given the sky map and the orange dotted arrows the sky map given the power spectrum. As shown, the non-centered Gibbs sampler explores the marginals of the high signal-to-noise coefficients much less efficiently, left panel, than those of the low signal-to-noise ones, right panel.}\label{fig:NonCenteredMoves}
\end{center}
\end{figure}

The second problem of the non-centered Gibbs sampler is related to the sampling of the power spectrum conditionally on the observed data and the signal map as discussed in, e.g.,~\cite{jewell_markov_2009-2}. Indeed, when looking at the distribution in equation~\eqref{NCPS} we see that for the low signal-to-noise ratio parameters, we can make large moves in the parameters space and the value of the density will not change much because the noise is much bigger. Unfortunately the opposite is true for high signal-to-noise ratio parameters: when the noise is small compared to the power spectrum, making large moves will make large changes in the value of the density, leading to a small acceptance rate in the Metropolis-Hasting algorithm. \textcolor{black}{ This is in addition to the fact that a mere use of the non centered parametrization already worsens the mixing properties of the Gibbs sampler of the marginal distributions for the high signal-to-noise ratio coefficients as visualized in Figure~\ref{fig:NonCenteredMoves}}. This intuition is confirmed by the experiments made in~\cite{jewell_markov_2009-2}. 

Consequently, we still need to find an alternative algorithm that is capable of sampling efficiently the high and low SNR simultaneously.

\section{Interweaving\label{interweaving}}
\subsection{Algorithm}
The idea of interweaving, also called ASIS in the statistical literature, see~\cite{yu_center_2011}, allows us to capitalize on the properties of the centered and non-centered Gibbs algorithm presented earlier. This is done by combining both these samplers together rather than simply alternating between them. In this section we apply this idea to the power spectrum estimation.

The interweaving scheme proposed here applies first the Gibbs kernel as described in Section~\ref{CenteredGibbs}, followed by changing the variable to get a non-centered version of the algorithm and finally concluding by sampling the power spectrum as explained in Section~\ref{NonCenteredGibbs}. These steps are implemented in Algorithm~\ref{AlgoASIS}.
\begin{algorithm}\label{AlgoASIS}
\caption{Iteration $t$ of ASIS}
    \KwInput{$(\{C_{\ell}\}_{t}, s_{t})$}
    \KwOutput{$(\{C_{\ell}\}_{t+1}, s_{t+1})$}
    $s_{t+0.5} \sim p(s|d, \{C_{\ell}\}_{t})$\\
    $\{C_{\ell}\}_{t+0.5} \sim p(\{C_{\ell}\}|s_{t+0.5})$\\

    $\tilde{s}_{t+0.5} = \mathbf{C}(\{C_{l}\}_{t+0.5})^{-1/2}s_{t+0.5}$\\

    $\{C_{\ell}\}_{t+1} \sim p(\{C_{\ell}\}|d, \tilde{s}_{t+0.5})$\\
    $s_{t+1} =C(\{C_{\ell}\}_{t+1})^{1/2}\tilde{s}_{t+0.5}$
\end{algorithm}
The first two steps of the algorithm are the usual centered Gibbs sampler, Section~\ref{CenteredGibbs}. The third step constitutes a change of variable that shifts to the non-centered version of the Gibbs sampler. The fourth step effectively samples the power spectrum from the non-centered parametrization, while the fifth goes back to the centered one.

We note that we can look at the interweaving algorithm as an Alternating Subspace-Spanning Resampling algorithm (ASSR), see \cite{liu_alternating_2003} with the underlying MCMC algorithm being the centered Gibbs sampler and the mapping defined as $\mathcal{M}(\{C_{\ell}\}, s) = (\{C_{\ell}\}, \tilde{s})$.

Intuitively, the algorithm will have better mixing properties than the centered and non-centered Gibbs sampler algorithms. First, interweaving will mix as well as the centered Gibbs sampler on the marginals of the high SNR parameters, thanks to Steps 1 and 2, see Section~\ref{CenteredGibbs}. It will also mix as well as the non-centered Gibbs sampler on the marginals of the low SNR ones, thanks to the change of variable and sampling in Steps 3 and 4. Second, we are not only exploiting the strength of each algorithm. We can expect interweaving to show a ``compound effect'': in the case of the high SNR parameters will still benefit a bit from the non-centered step, however inefficient it may be, and vice-versa.

So far we have proposed an algorithm that we expect to behave nicely on a broad range of signal-to-noise ratios. But the constrained realization step is still a problem: whatever the mixing properties of the algorithm we are using, the cost of one iteration is still very high and this is expected to continue to be a major hindrance for the applications.

\section{Constrained realization step}
\label{CRStep}
Solving the constrained realization equation, equation~\eqref{eq:PCG}, is a problem for several reasons. First, this system is highly dimensional and dense and computing explicitly the inverse of its system matrix, $Q$, would be really time consuming, not to mention the memory requirements to store it. These issues can be efficiently handled by the use of an iterative solver, most commonly a preconditioned conjugate gradient (PCG) algorithm. However, iterative algorithms solve the system only up to some pre-defined accuracy and require sometimes a large number of iterations to provide a sufficiently precise solution. Trading on this may speed up time to solution but can result in a bias, effects of which are hard to quantify.

{\color{black}
One solution would be to add a Metropolis-Hastings step after we proposed a new sky map sample using equation~(\ref{eq:PCG}). This indeed would ensure that the accepted constrained realization solutions conform indeed with the desired posterior.
However, such a naive implementation would lower the acceptance rate resulting into high autocorrelations between the successive samples.
}

We propose two alternative solutions to these two problems in this section.
First, we present an auxiliary variable scheme that allows us to add a Metropolis-Hastings step without reducing the efficiency of the sampler. Second, we introduce another auxiliary variable that allows us to eliminate altogether any need to sample exactly from a high dimensional normal distribution with a dense covariance matrix.

These two algorithms leave distribution in equation~(\ref{eq:constrainredRealConditional}) invariant, leading therefore to a valid Gibbs Sampler.

\subsection{Reversible jump perturbation optimisation step}
\label{RJPO}
{\color{black}
Our first approach is based on an algorithm called in the statistical literature the Reversible-Jump Perturbation Optimization (RJPO) algorithm described in~\cite{gilavert_efficient_2015}.
}

We start from augmenting the model with an auxiliary
variable $z$ such that
\begin{equation}\label{auxVar}
z|s \sim \mathcal{N}(\Q s + \Q\mu, \Q)
\end{equation}
where $\Q$, $\mu$ are defined equation~(\ref{eqn:constRealPars}) and $s$ is distributed according to equation~(\ref{eq:constrainredRealConditional}).
We then perform a Metropolis-Hastings move on this augmented target.

Our proposal consists in the following deterministic, differentiable and reversible transformation:
\begin{equation}\label{deterministic}
    \phi(s,z) = ( -s + f(z), z)= (s', z).
\end{equation}
Following~\cite{gilavert_efficient_2015}, the Metropolis acceptance rate for this proposal writes:
\begin{equation}\label{MH_ratio}
\min (1, e^{-r(z)^{t} (s - s')}),
\end{equation}
where  $r(z) \eqdef z - \Q f(z)$.

On choosing $f(z) = \Q^{-1}z$, the acceptance rate of the Metropolis-Hastings scheme is one, and we accept every proposed move. In addition, as shown in~\cite{gilavert_efficient_2015}, this choice of $f(z)$ leads to uncorrelated successive samples.

Note that in this case $s'= - s + \Q^{-1} z = - s + \Q^{-1}(\Q s + \eta) =  \Q^{-1}\eta$ where $\eta \sim \mathcal{N}(\Q\mu, \Q)$ which means we are solving the exact same system as for sampling from equation~(\ref{eq:constrainredRealConditional}) in the usual centered Gibbs sampler.

As we explained before, the problem is that we are unable to solve this system exactly and instead we have to relay on some iterative algorithms that in the interest of time we stop once some predefined precision has been reached. The scheme considered here allows to account on such effects.

Indeed, instead of defining $f(z) = \Q^{-1}z$, we can define it as the output of a truncated iterative solver like the preconditioned conjugate gradient algorithm. applied to the system $\Q f(z) = z$.  The acceptance rate of the corresponding  Metropolis-Hastings algorithm will not be 1 and the correlations between two successive samples will not be 0 anymore. Since the ratio depends on $r(z)$, the more precisely we solve the system, the higher the acceptance rate, but so is the computational cost. With this algorithm, we are facing a computational efficiency/autocorrelation tradeoff.

Let us denote $\hat{u}$ the approximate solution of $\Q f(z) = z$. Now we have $s' = - s + \hat{u}$ and $r(z) = z - \Q\hat{u} = z - \Q( s + s') =  \eta - \Q s'$. Finally the algorithm reads as Algorithm~\ref{AlgoRJPO}.
\begin{algorithm}\label{AlgoRJPO}
\caption{RJPO algorithm}
    Sample $\eta \sim \mathcal{N}(\Q\mu, \Q)$\\
    Solve $\Q\hat{s} = \eta$ approximately\\
    Compute $\alpha = \min (1, e^{-r(z)^{t} (s - \hat{s})})$ where $r(z) \eqdef  \eta - \Q\hat{s}$.\\
    With probability $\alpha$, set $s' = \hat{s}$, otherwise set $s' = s$.
\end{algorithm}

It is remarkable that the first and second steps of this RJPO algorithm are exactly the same system as for sampling from distribution in equation~(\ref{eq:constrainredRealConditional}). We are just adding a Metropolis-Hastings step to ensure that we are leaving this distribution
invariant, however approximately we solve the system.

The presence of such a Metropolis step allows us to solve the system with an arbitrary precision without biasing the Metropolis-within-Gibbs algorithm. Thus, we can spare some computation time by decreasing the precision required to solve the system.

If one chooses to solve the system exactly, the RJPO algorithm always accept the proposed move and the successive samples are uncorrelated. In this case, RJPO is exactly the same as the PO algorithm used to sample from equation~(\ref{eq:constrainredRealConditional}) in Section~\ref{CRSection}.

If instead we decide to solve the system only approximately, we introduce correlations between successive samples and the acceptance rate will depend on how approximate we solve it: the more precise we are, the higher the acceptance rate.

Even though the RJPO algorithm has nice properties, it still involves solving a very high dimensional system, at least very approximately. We also have to arbitrate between a lower computing time and higher autocorrelations. In the next section we present another auxiliary variable scheme that completely bypasses such inconveniences.

\subsection{Augmented Gibbs step}
\label{augGibbs}
Instead of shortening the computing time needed to solve the constrained realization linear system as we did in the previous subsection, we may avoid it completely and rely on a different MCMC scheme. The dimension is huge though, and we would like to avoid the computation of an acceptance ratio. A Gibbs sampler seems a natural solution. 
{\color{black}
The relevant algorithm has been originally proposed in the statistical literature in~\cite{Marnissi2018}. In this Section we describe it and adapt the generic algorithm to the specific case of the CMB power spectrum estimation.}

\subsubsection{Gibbs step}
Instead of sampling directly from the conditional distribution, equation~(\ref{eq:constrainredRealConditional}):
\[ \pi(s | \{C_{\ell}\}, d) \]
we augment it with an auxiliary variable $v$ so that sampling from $\mathcal{L}(v|s, \{C_{\ell}\}, d)$ and $\mathcal{L}(s|v, \{C_{\ell}\}, d)$ is easier. We choose a $v$ such that:
\begin{equation}\label{eq:augmentedGibbsV}
 v | s, \{C_{\ell}\}, d \sim  \mathcal{N}(\Gam \Ytilde s, \Gam)
\end{equation}
where $\Gam \eqdef (\beta \boldsymbol{I} - \NoiseMatrix^{-1})$ and $\beta$ is a scalar chosen so that $\Gam$ is positive definite. 
This gives us the following conditional distribution (up to an irrelevant prior on $v$),
\begin{equation}\label{eq:augmentedGibbsS}
 s | v, \{C_{\ell}\}, d) \sim  \mathcal{N}(\M \Ytilde^{T}(v + \NoiseMatrix^{-1}d), \M )
\end{equation}
where $\Beam$ is the beam matrix and $\M \eqdef (\dfrac{\beta \Npix}{4\pi}\Beam^{2} + \C^{-1})^{-1}$. Note that both $\Gam$ and $\M$ are diagonal -- or block diagonal matrices in the case of temperature and polarization.
We note that we can sample efficiently from these two conditional distributions, and consequently we are able to sample from the distribution in equation~(\ref{eq:constrainredRealConditional}) as well and to do so without any need for solving the constrained realization problem. Indeed, we can simply use the Gibbs sampler and draw pairs of $(s,v)$ consecutively from their conditional distributions and
since $\int \pi(s,v|d, \{C_{\ell}\}) dv = \pi(s |d, \{C_{\ell}\})$, we merely discard $v$ at the end. Such a scheme leaves distribution in equation~(\ref{eq:constrainredRealConditional}) invariant.

Even though this augmented Gibbs step is computationally efficient, its overall performance will mainly depend on the correlations between $v$ and $s$. If we write the joint distribution of $(s,v)$, we realize that they are jointly Gaussian with covariance matrix,
\begin{equation}\label{covMatrix}
\left[ 
\begin{array}{c|c} 
  \Sig & \Sig \Ytilde^{T} \Gam \\ 
  \hline
  \Gam \Ytilde \Sig & \Gam + \Gam \Ytilde^{T}\Sig\Ytilde\Gam 
\end{array} 
\right] 
\end{equation}
Looking at equation~(\ref{covMatrix}), we see that $\Sig$, defined in equation~(\ref{eq:constrainredRealConditional}), is influencing the correlations between $s$ and $v$. 
From our earlier analysis, see also~\cite{eriksen_power_2004}, 
this may be a problem since $\Sig$ may be dense for the high signal-to-noise ratio parameters. We can expect this Gibbs step to show poor mixing on their marginals. However, for lower SNR ratio parameters, $\Sig$ tend to be band diagonal and we can expect this Gibbs move to be much more efficient. We confirm this expectation with help of numerical experiments in Section~\ref{cutSkyPlanckMask}.

\subsection{Overrelaxation}
The overrelaxation method, see~\cite{overrelaxation} for the statistical background, is a way around these strong correlations.
Instead of sampling successively from distributions in equations~\eqref{eq:augmentedGibbsV} and~\eqref{eq:augmentedGibbsS}, we are going to sample from,
\begin{equation}\label{eq:overrelaxV}
    v^{t+1} = \Gam \Ytilde s^{t} + \gamma (v^{t} - \Gam \Ytilde s^{t}) + \Gam^{1/2}(1 - \gamma^{2})^{1/2} Z_{1}
\end{equation}
and,
\begin{multline}\label{eq:overrelaxS}
    s^{t+1} = \M\Ytilde ^{T}(v^{t+1} + \NoiseMatrix^{-1}d) + \gamma (s^{t} - \M\Ytilde ^{T}(v^{t+1} + \NoiseMatrix^{-1}d)) \\+ \M^{1/2}(1 - \gamma^{2})^{1/2} Z_{2}
\end{multline}
where $Z_{1}, Z_{2}$ are two independent standard normal variables, $Z_{1}$ has the dimension of $v$ and $Z_{2}$ of $s$. Here $\gamma\in ]-1, 1[$ is a parameter chosen by the user.\\
It is straightforward to show that the move in equation~(\ref{eq:overrelaxV}) {\color{black} samples} the distribution in equation~(\ref{eq:augmentedGibbsV}), i.e., the distribution of $v_{t+1}$ is given by equation~(\ref{eq:augmentedGibbsV}) if that of $v_t$ is, and that the move in equation~(\ref{eq:overrelaxS}) leaves the distribution in equation~(\ref{eq:augmentedGibbsS}) invariant. In addition, it has been argued in the statistical literature~\cite{overrelaxation} that such a "symmetrical" conditional move around the mean make it possible for the Gibbs sampler to move in a consistent direction in the presence of correlations, thus suppressing the random-walk behavior of the Gibbs sampler.

\section{Experiments}\label{Experiments}
In this section we consider several experiments.  For the first comparison of our algorithms, we assume that we observe the entire sky. This way, the covariance matrices are diagonal, the centered, non-centered and interweaving algorithms are computationally cheap and we can easily draw many samples.

In the second round of experiments we assume exactly the same setting as the first one, except that we apply the 80\% Planck mask leading to a posteriori coupled multipoles. \textcolor{black}{All the masks and products of the Planck mission can be found on their website \footnote{http://pla.esac.esa.int/pla/home}}

In both cases in order to test our algorithms in the circumstances reflecting potential future applications we assume noise levels and the resolution reflecting roughly those of the future CMB satellite mission, LiteBird~\cite{LiteBIRD}.

The final set of experiments is designed to mimic a ground-based setup. We take here very roughly the parameters of the 90GHz frequency channel of the Simons Observatory, see~\cite{Simon}. We assume a sky coverage of 37\%, what leads to even more strongly coupled multipoles.

{\color{black}In all the cases the simulated maps contain the CMB signal and noise only.}

\subsection{Polarization full-sky experiment}
\label{fullskyExp}
For this first experiment comparing interweaving and the centered and non centered Gibbs algorithms, we assume we observe the entire sky and that the noise covariance matrix writes $N = \alpha^{2} I$, where $I$ is the identity matrix of dimension $\Npix$. For ease of implementation we are doing inference on $EE$ and $BB$
power spectra only, assuming that only maps of the $Q$ and $U$ Stokes parameters are available. In this case, owing to the full sky coverage, we can exactly sample a map from the constrained realization step in equation~(\ref{eq:constrainredRealConditional}) at no cost since the signal covariance matrix is diagonal in the harmonic domain. In addition, the power spectrum coefficients are a posteriori independent and we can derive an analytical expression for each marginal distribution.

Regarding the set-up, we choose $\mathrm{NSIDE} = 256$ with $\lmax = 512$ and we apply an instrumental beam of $30$-arcmin fwhm. We choose a rms noise of $\alpha = 0.2\mu K$-arcmin. Since the BB spectrum coefficients have a very low signal-to-noise ratio for the highest multipoles, we recover the BB spectrum with  bins progressively becoming wider bins starting at $l=396$, corresponding to $\SNR= 0.24$, and get a total of $412$ spectrum amplitudes instead of the $512$ initial ones.

We consider in this case three algorithms: Centered -- corresponding to the Centered Gibbs, Sect.~\ref{CenteredGibbs}, non-Centered -- corresponding to the non-Centered Gibbs, Sect.~\ref{NonCenteredGibbs}, and ASIS, Sect.~\ref{interweaving}. For the two last algorithms, we first run $10$ chains of length of $300$ samples for each algorithm and use these samples to tune the proposal distributions of the non-centered power sampling step. Once we have the covariance matrices of the proposal distributions, we run $10$ chains of $10^{4}$ iterations for both non-Centered and ASIS algorithms and compute the relevant metrics on this basis. For the Centered algorithm we run $10$ chains of $10^{4}$ iterations.
For each of the algorithms we then compute the so-called Integrated Autocorrelation Time ($\IAT$)  as a function of the multiple number, where $\IAT$ is defined as,
\[\IAT_{\ell} \eqdef 1 + 2 \sum_{k=1}^{N_{\mathrm{lag}}} \rho_{k}^{\ell}\]
and $\rho_{k}^{\ell}$ is the autocorrelation of the chain at lag $k$ for $C_{\ell}$ defined at stationarity as:
\begin{equation}
    \rho_{k}^{\ell} \eqdef \dfrac{\Cov(C_{\ell,0}, C_{\ell,k})}{\Var(C_{\ell}|d)}
\end{equation}
with $C_{\ell,0}\sim p(C_{\ell}|d)$ and $C_{\ell,k}$ is obtained after $k$ iterations of Gibbs sampler, starting at $C_{\ell,0}$.

\begin{figure*}
\centering
\includegraphics[scale=0.6]{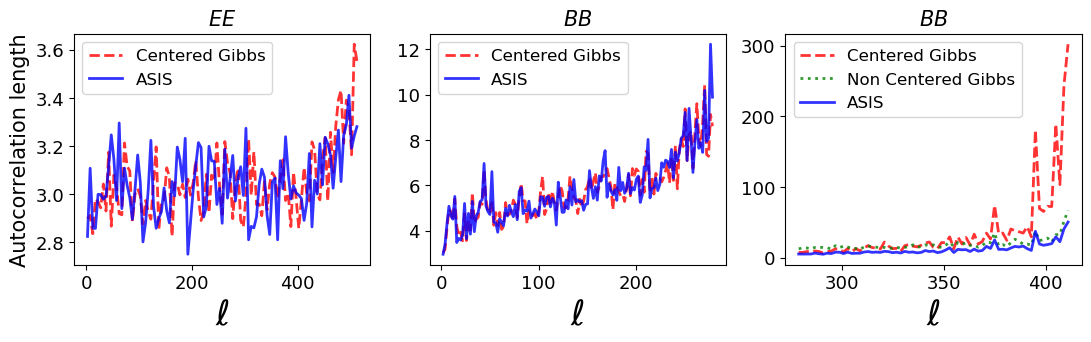}
\caption{Integrated autocorrelation times against multipole for each algorithm and the full-sky case. The results for the marginals of the BB coefficients are split into two graphs to show better the details of the observed behavior. \label{fig:autoLengthsEllFullSky}
}
\end{figure*}

\begin{figure*}
\centering
\includegraphics[scale=0.39]{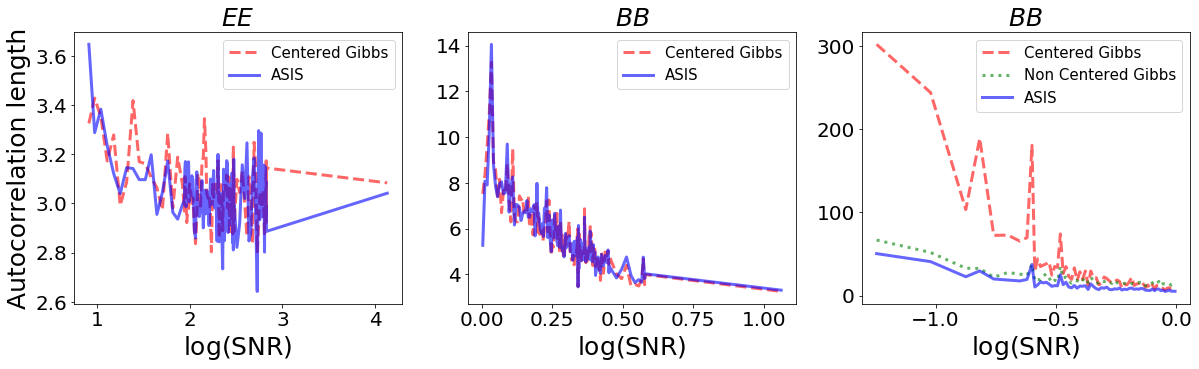}
\caption{Integrated autocorrelation time against log(SNR) for each algorithm and the full-sky case. The results for the marginals of the BB coefficients are split into two graphs to show better the details of the observed behavior. }\label{fig:autoLengthsSNRFullSky}
\end{figure*}

Figures~\ref{fig:autoLengthsEllFullSky} and~\ref{fig:autoLengthsSNRFullSky} show the IAT for each algorithm against the multipole number and the logarithm of the signal-to-noise ratio respectively. Since it is known~\cite{jewell_markov_2009-2} that the non-centered Gibbs sampler does not mix well -- however good is our tuning -- for medium to high SNR, we only tuned it for coefficients with expected signal to noise ratio inferior to 1. For readability, we only show the results for these coefficients and we display the $BB$ case in two different panels to cover the entire range of the corresponding values. As expected, the interweaving ensures that the ASIS algorithm mixes as well as the Centered one on signal-to-noise ratio superior to 1. However, when the SNR starts to be low, the integrated autocorrelation times of the centered Gibbs algorithm increase sharply compared to those of the interweaving. Note also that for the lowest signal-to-noise ratios, the non-Centered  sampler performs better than the centered version, as expected, and that ASIS has even lower integrated autocorrelation times than non centered Gibbs. These results are in agreement with the analysis of Section~\ref{interweaving}.

\begin{figure}
\centering
\includegraphics[scale=0.27]{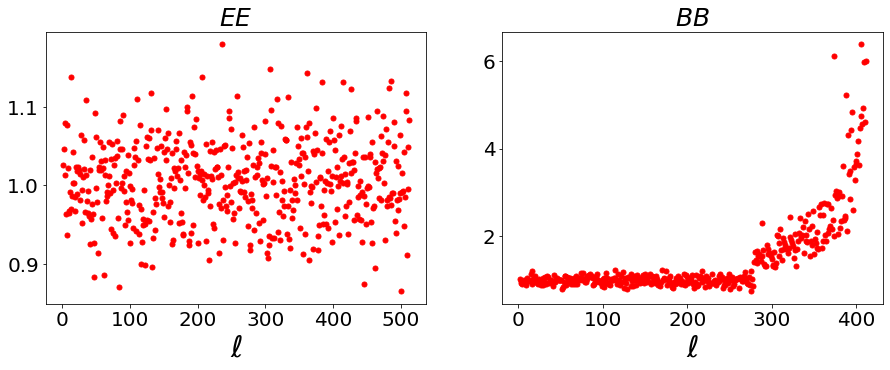}
\caption{Ratios of Integrated Autocorrelation Time against the multipole number for the centered Gibbs, numerator, and the ASIS, denominator, algorithms and the full-sky case. A ratio larger than 1 indicates that centered Gibbs is performing better than interweaving in terms of autocorrelation time.}
\label{fig:ratiosCentered}
\end{figure}

\begin{figure}
\centering
\includegraphics[scale=0.21]{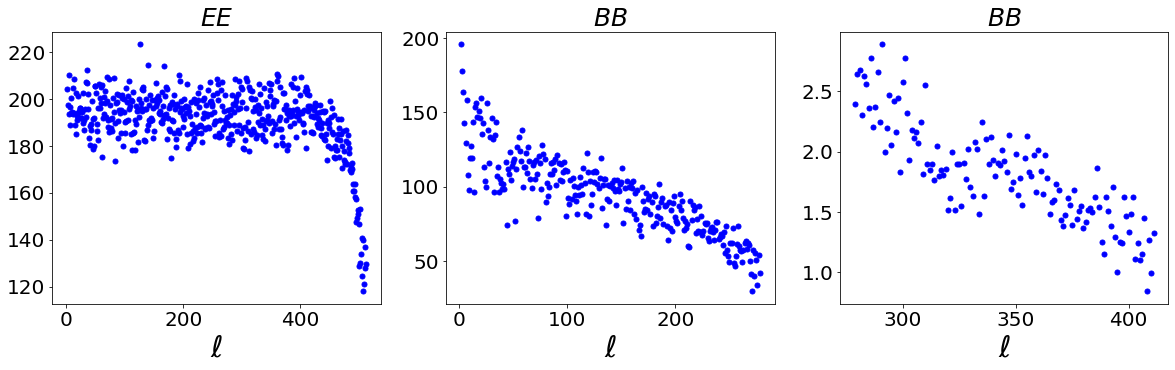}
\caption{As Fig~\ref{fig:ratiosCentered} but for the non centered Gibbs algorithm instead of the centered one.}
\label{fig:ratiosNonCentered}
\end{figure}

\begin{figure}
\centering
\includegraphics[scale=0.27]{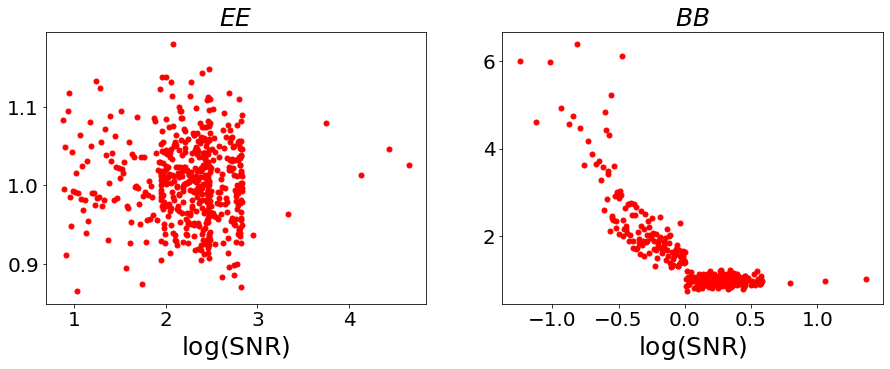}
\caption{Ratios of Integrated Autocorrelation Time against the logarithm of the signal-to-noise ratio for the centered Gibbs, numerator, and the ASIS, denominator, algorithms. The full-sky case.}
\label{fig:ratiosCenteredSNR}
\end{figure}

\begin{figure}[htpb]
\centering
\includegraphics[scale=0.21]{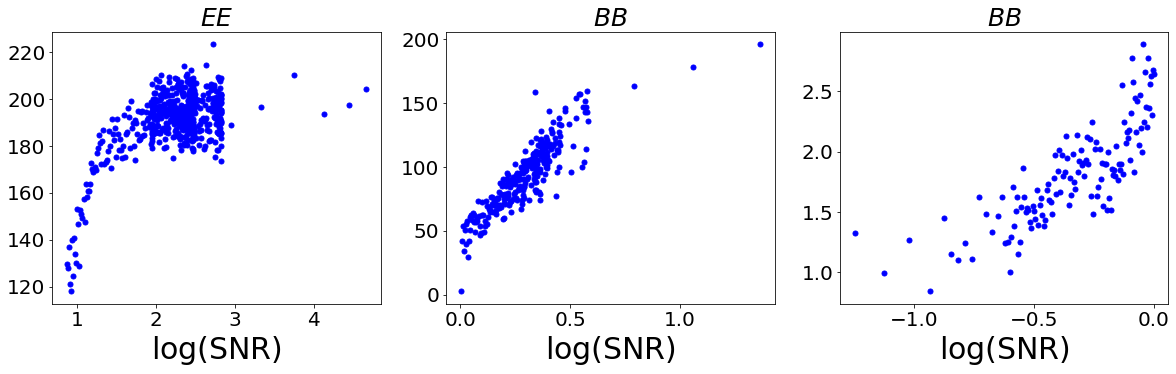}
\caption{Ratios of Integrated Autocorrelation Time against signal-to-noise ratio for non centered Gibbs on interweaving. Note that, for readability, the results for the BB coefficients are split in two graphs, given the broad range of values they span. The full-sky case.}
\label{fig:ratiosNonCenteredSNR}
\end{figure}

In order to have a clearer picture of the respective performances of the algorithms, we plot the ratios of $\IAT^{\mathrm{centered}}/\IAT^{\mathrm{asis}}$ and $\IAT^{\mathrm{non centered}}/\IAT^{\mathrm{asis}}$ against the multipoles in Figures~\ref{fig:ratiosCentered} and~\ref{fig:ratiosNonCentered}, respectively. We show them against the log signal-to-noise ratio in Figures~\ref{fig:ratiosCenteredSNR} and~\ref{fig:ratiosNonCenteredSNR}. It is clear from these plots that the interweaving algorithm inherits the excellent mixing properties of the centered Gibbs on the marginal distributions of the high signal-to-noise ratio parameters while outperforming it on the ones with lower signal-to-noise ratios. In addition, the interweaving algorithm outperforms the non-centered Gibbs one over the lower signal-to-noise range. We provide example of histograms for a wide range of signal-to-noise ratios in Appendix~\ref{full-sky appendix}.

This simple experiment shows how good the mixing properties of the interweaving algorithm are on the full range of multipoles as it is able to sample efficiently for high and low signal-to-noise ratios.  Our analysis of the respective efficiencies of the algorithms does not take into account the computing time. However, for the full sky cases, these algorithms are very cheap computationally. In the presence of a sky-mask, the computational time becomes an issue. Moreover, the power spectrum coefficients, $\{C_{\ell}\}$, are no longer independent and this may hinder the non-centered power spectrum sampling step.
In order to test the algorithms in more realistic contexts, we consider a second experiment with a cut-sky in the next section.

In the following, we will also test a broader set of the plausible algorithms building on the ideas introduced earlier. Specifically, in addition to the ``Centered'' and ``ASIS'' algorithms, we will consider an algorithm called ``ASIS RJPO''  combining the interweaving with the RJPO step, respectively, Sect.~\ref{RJPO}. We will also use a number of algorithms based on either the Centered or the ASIS ones, but which use some number of the augmented Gibbs sampler iterations, Sect.~\ref{augGibbs}, to replace the need for the PCG solution.  In these cases we refer to these algorithms by the name of the initial algorithm followed by an integer defining the number of the augmented Gibbs iterations. For example, ``Centered 1'' is the usual centered Gibbs algorithm with the PCG solver replaced by a single iteration of the augmented Gibbs sampler. In these cases, we set $\beta= \alpha^{-2} + 10^{-14}$, where $\alpha^2$ defines the noise level. \textcolor{black}{For a more general noise covariance matrix, this should be generalized and we should choose a $\beta$ that is greater than the largest eigenvalue of the inverse of this matrix.}
We also consider a algorithm called ``Centered overrelax'', which is the centered Gibbs algorithm with the PCG step replaced by two iterations of overrelaxation plus one iteration of classical augmented Gibbs sampler. In these cases, we set  $\gamma = -0.995$ chosen to be close to $-1$ to deal with the strong correlations of the covariance matrix, equation~\eqref{covMatrix}. Table~\ref{algoTable} summarizes all these algorithms.

\subsection{Nearly full-sky polarization experiment}
\label{cutSkyPlanckMask}
The set-up of this experiment is exactly the same as the one Section~\ref{fullskyExp} except that we apply the 80\% Planck sky mask, that we plot in Appendix~\ref{subsec:masks}. We use the same binning and blocking schemes as in the previous section.

Because we do not analyze the full sky, we cannot access the true posterior distribution and the system to solve for the constrained realization step is no longer diagonal. We therefore have to rely on the TPO algorithm and the PCG solver~\cite{gilavert_efficient_2015} as done in previous works ~\cite{jewell_markov_2009-2}, \cite{eriksen_power_2004} and explained in Section~\ref{CRSection} or rely on the augmented Gibbs sampler.

In the algorithms employing the PCG solver we follow \cite{eriksen_power_2004} and set the error threshold to $10^{-6}$, except for ASIS RJPO, where it is set to $10^{-5}$. In all cases we use a diagonal preconditioner.

\begin{center}
\begin{table*}
\advance\leftskip-0.5cm
\begin{tabular}{ |c|c|c|  } 
 \hline
 \textbf{Algorithm} & \textbf{Constrained realization} & \textbf{power spectrum sampling} \\ [1ex] 
 \hline 
 Centered & PCG & centered move \\ [1ex]
 ASIS & PCG & centered + non-centered moves \\ [1ex]
 ASIS RJPO & RJPO & centered + non-centered moves\\ [1ex]
 ASIS 1 & auxiliary variable & centered + non-centered moves\\ [1ex]
 ASIS 20 & auxiliary variable & centered + non-centered moves \\ [1ex]
 ASIS 65 & auxiliary variable & centred + non-centered moves \\ [1ex]
 Centered 1 & auxiliary variable & centered move \\ [1ex]
 Centered overrelax & overrelaxation & centered move\\
 
 \hline
\end{tabular}
\caption{\label{algoTable} 
Summary of the algorithms used in the experiments described in the text highlighting the approaches they implement to address the constrained realization and the power spectrum sampling steps.}
\end{table*}
\end{center}
We tune the algorithms as follows. We run each one of them for a few hundreds iterations. Based on the results, we estimate the covariances of the marginal of each multipole and use them as the proposal covariances -- multiplied by a scalar inferior to one -- for the actual run, targeting a $25$\% acceptance rate. After tuning, every algorithm is run for $10^{3}$ iterations, except for Centered overrelax and Centered 1 which are run for $10^{5}$ iterations since they are computationally cheaper.

We first look at the Effective Sample Sizes (ESS) per second of each algorithm. If we run the algorithm for $N$ iterations, the ESS for the marginal of a power spectrum coefficient with multipole $\ell$ is defined as:
\[ \ESS_{\ell} \eqdef \dfrac{N}{\IAT_{\ell}}\]
where ${\IAT_{\ell}}$ is defined in Section~\ref{fullskyExp}. The $\ESS$ per second, for each marginal, is then defined as the $\ESS$ for the $N$ iterations divided by the CPU time in second needed for the $N$ iterations. Obviously, for any $\ell$, the greater $\ESS_{\ell}$ per second, the better.

\begin{figure*}[htpb]
\centering
\includegraphics[scale=0.45]{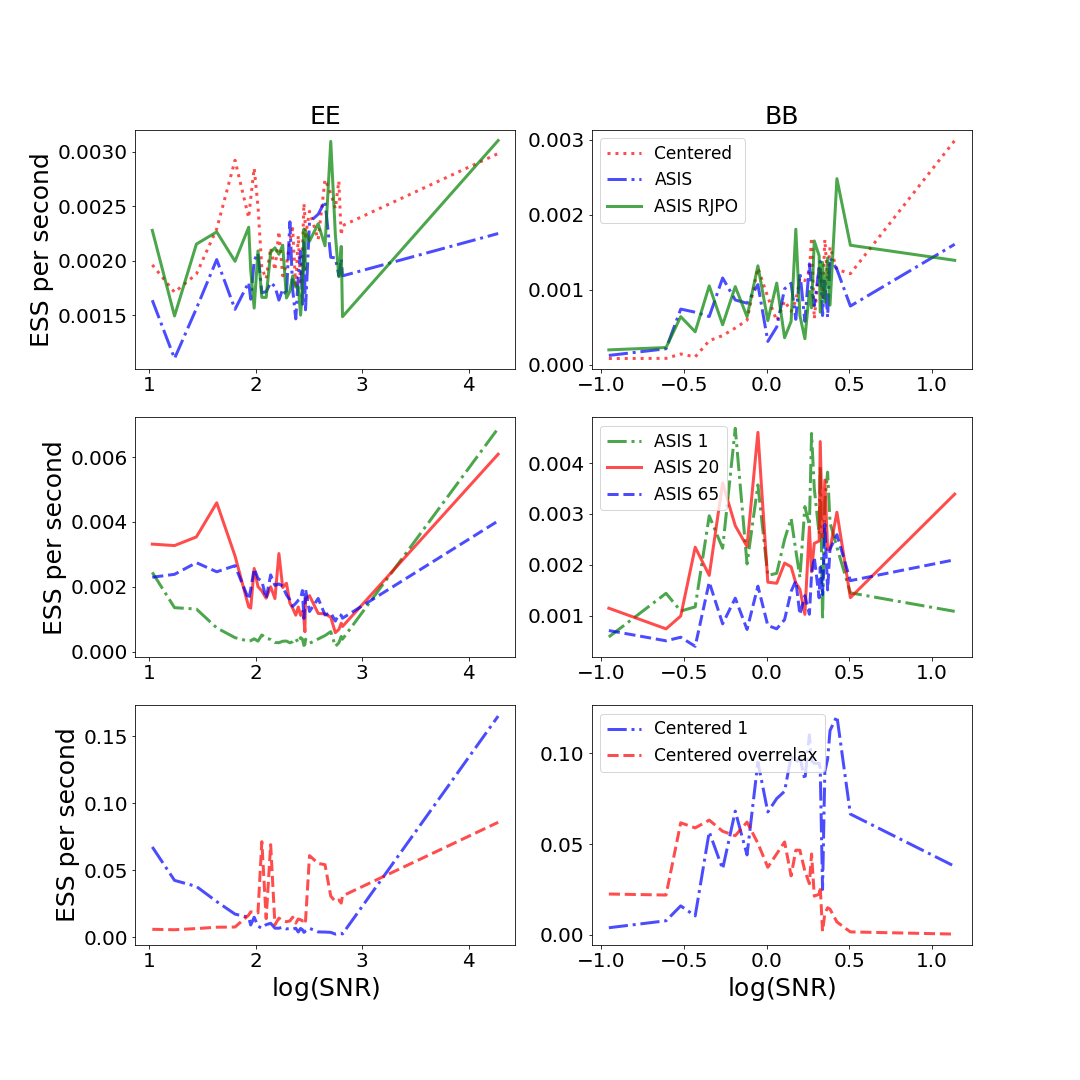}
\caption{Effective Sample Size per second against $\log(\SNR )$ for each algorithm. For the sake of clarity, we group the algorithms with a similar performance and plot different groups in separate panels. The left columns shows the results for the $EE$ and the right one for the $BB$ spectra. The case of nearly full-sky coverage.
}
\label{fig:EssSecSNR}
\end{figure*}

We plot the $\ESS$ per second in Figure~\ref{fig:EssSecSNR}. Centered 1 and Centered overrelax outperform the other algorithms in term of $\ESS$ per second, whatever the SNR. Otherwise, the algorithms using the PCG sampling step seem to be performing better on $EE$ coefficients than the ones using an auxiliary Gibbs step. This is especially the case of ASIS 1 which clearly underperforms. The opposite is true for $BB$. We can easily explain these observations: the augmented Gibbs constrained realization step is much cheaper than a PCG resolution of the system, but it also leads to much worse mixing properties on $EE$ but not on $BB$. We are facing a trade off between computing time and mixing on $EE$: the smaller the number of augmented Gibbs step, the faster the algorithm but the greater the autocorrelations. The same holds for ASIS RJPO. We must then find the number of Gibbs steps maximizing the $\ESS$ per second. But overall it seems these algorithms will not perform as well as their PCG counterparts on $EE$.

The reader should also note that the non-centered step of interweaving comes with a cost that cannot be reduced: in our case roughly $130$ spherical harmonic synthesis operations. That is why ASIS 1 performs much worse than ASIS 20 and ASIS 65: as the algorithm has to perform at least $130$ spherical harmonic transforms, one could as well do $20$ augmented Gibbs constrained realization step instead of 1, improving the mixing properties of the algorithm without increasing the computing time so much, leading to a better $\ESS$ per second. 

The picture is different for the $BB$ spectrum coefficients: the augmented Gibbs constrained realization step is mixing much better given their lower signal-to-noise ratios and hence the algorithms using such a step have a better $\ESS$ per second than their PCG counterparts. Note that Centered 1 and Centered overrelax are outperforming all the other algorithms by far. That is because it comes at almost no cost -- only one spherical harmonic analysis and one synthesis per iteration. This has to be compared to the heavy cost of the PCG solver, typically $150$ PCG iterations, complemented by $2$ spherical harmonics transform per iteration, of the Centered, ASIS and ASIS RJPO algorithms, and to the incompressible cost of the non centered step of the ASIS 1, ASIS 20 and ASIS 65 algorithms. Since the augmented Gibbs step mixes well on this SNR, Centered 1 and Centered overrelax are good mixing and cheap algorithms, hence their $\ESS$ per second is much better. One must be careful though: this behavior tends to fade on very low SNR: the centered parametrization has greater and greater autocorrelations as the SNR decreases.

In order to get a better idea of the relative performances of the algorithms, we examine the ratios of $\ESS$ per second of each algorithm on the $\ESS$ per second of the usual centered Gibbs and of the interweaving algorithm, Tables~\ref{TableCenteredESSPlanckEE} and \ref{TableCenteredESSPlanckBB}. 

These tables confirm the behavior we described above: on $EE$ coefficients ASIS 1 and ASIS 65 are outperformed by Centered and ASIS in terms of $\ESS$ per second, while ASIS 20 seems to perform similarly. Note that the algorithms tend to perform worse in comparison to Centered than compared to ASIS: thought ASIS and Centered have roughly the same mixing on $EE$, ASIS is more expensive than Centered. In addition, Centered is outperforming ASIS on $EE$ because it is a bit cheaper. On $BB$ however, each algorithm seems to  outperform Centered. Again, this is because the augmented Gibbs constrained realization step leads to as good a mixing as the PCG resolution while dramatically reducing the overall cost of the algorithms, leading in turn to a much better $\ESS$ per second. As for ASIS, its $\ESS$ per second is greater than the one of Centered only for the lowest signal-to-noise ratio coefficients. This indicates that we could probably have applied the non-centered step on these parameters only: the algorithm would have been cheaper while still having good mixing properties for the low SNR cases, leading to a much better ESS per second, on both $EE$ and $BB$.

\begin{figure}[htpb]
    \centering
    \includegraphics[scale=0.24]{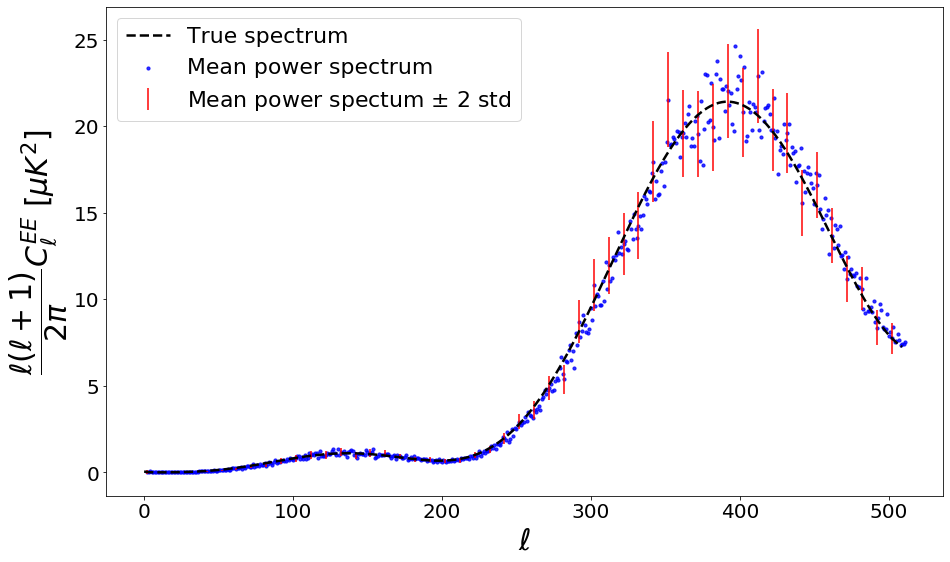}\\
    \vspace{1cm}
    \includegraphics[scale=0.24]{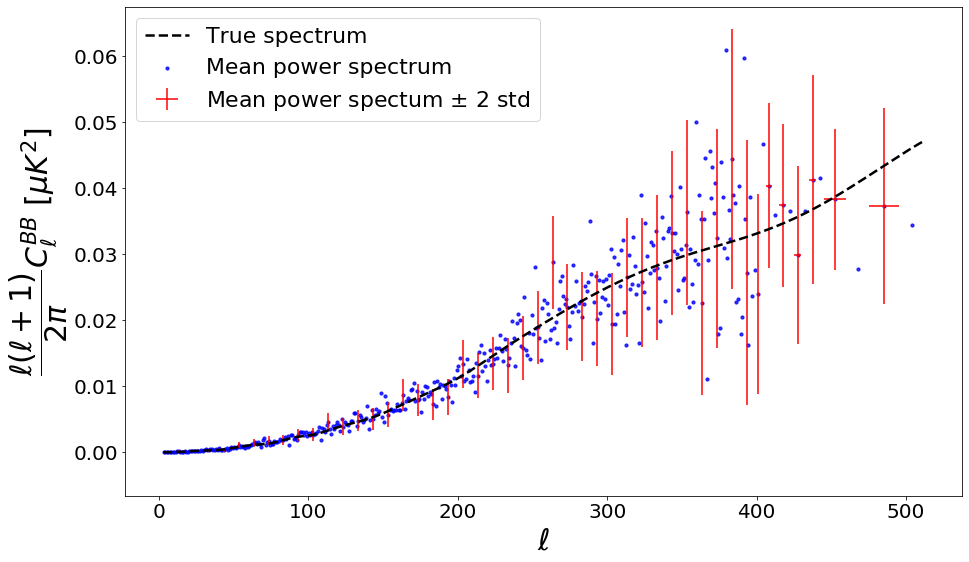}
    \caption{\textcolor{black}{Comparison of posterior power spectrum to the true power spectrum for the nearly full-sky experiment. The blue points  correspond to the mean power spectrum. The top and bottom of the horizontal bars of the crosses correspond to the mean plus/minus two standard deviation. The horizontal bars correspond to the $\ell$ range spanned by the binning scheme. We only plot the crosses for one every ten multipoles on the non binned part. On the binned part, we only plot the crosses for one every two multipoles. The panels show $EE$ (top) and $BB$ (bottom) power spectra.}}
    \label{fig:recoveredSpec}
\end{figure}

We should pay a closer attention to Centered 1 and Centered overrelax. These algorithms are computationally cheap compared to any other algorithm. Hence, whatever their mixing properties for the $EE$ coefficients and for the very low SNR parameters, their $\ESS$ per second is much higher. In addition, the $\ESS$ per second of Centered overrelax is higher than the one of Centered 1 on $EE$ coefficients, showing that these step is handling the strong correlations better.\\
Finally, Figure~\ref{fig:recoveredSpec} shows the empirical mean posterior of Centered overrelax with the two standard deviations intervals. The solid black lines denotes the true spectrum. The recovered spectrum seems to match the true spectrum well.

\begin{center}
\begin{table}
\begin{tabular}{ |c|c|c|c|c|c|  } 
 \hline
 Algorithm &$5$th&$25$th&$50$th&$75$th&$95$th \\ [2ex] 
 \hline 
 ASIS & 0.544 & 0.685 & 0.772 & 0.88 & 1.061 \\ [1ex]\hline
 ASIS 1 & 0.095 & 0.126 & 0.16 & 0.225 & 0.9\\ [1ex]\hline
 ASIS 20 & 0.287 & 0.477 & 0.697 & 1.044 & 1.997\\ [1ex]\hline
 ASIS 65 & 0.396 & 0.575 & 0.786 & 0.988 & 1.288\\ [1ex]\hline
 ASIS RJPO & 0.647 & 0.79 & 0.896 & 1.013 & 1.217\\ [1ex]\hline
 Centered 1 & 1.015 & 1.862 & 2.915 & 5.303 & 29.31\\ [1ex]\hline
 Centered overrelax & 2.843 & 4.708 & 6.925 & 11.265 & 28.635\\ [1ex]\hline
 
\end{tabular}
\caption{\label{TableCenteredESSPlanckEE} 
For each algorithm, percentiles of their $\ESS$ per second relative to the $\ESS$ per second of the Centered algorithm for the $EE$ spectrum coefficients and the nearly full-sky experiment, see Section~\ref{cutSkyPlanckMask}.}
\end{table}
\end{center}

\begin{center}
\begin{table}
\begin{tabular}{ |c|c|c|c|c|c|  } 
 \hline
 Algorithm &$5$th&$25$th&$50$th&$75$th&$95$th \\ [2ex] 
 \hline 
 ASIS & 0.376 & 0.697 & 1.016 & 1.625 & 3.966 \\ [1ex]\hline
 ASIS 1 & 1.04 & 1.694 & 2.696 & 4.936 & 14.132\\ [1ex]\hline
 ASIS 20 & 
1.115 & 1.689 & 2.466 & 3.972 & 10.15\\ [1ex]\hline
 ASIS 65 & 0.567 & 0.967 & 1.477 & 2.522 & 6.103\\ [1ex]\hline
 ASIS RJPO & 0.391 & 0.644 & 1.007 & 1.719 & 4.097\\ [1ex]\hline
 Centered 1 & 
38.62 & 63.838 & 83.914 & 116.926 & 169.758\\ [1ex]\hline
 Centered overrelax & 
2.173 & 14.331 & 36.227 & 100.185 & 492.866\\ [1ex]\hline
 
\end{tabular}
\caption{\label{TableCenteredESSPlanckBB} 
For each algorithm, percentiles of their $\ESS$ per second relative to the $\ESS$ per second of the Centered algorithm for the $BB$ coefficients and the nearly full-sky experiment, see Section~\ref{cutSkyPlanckMask}.}
\end{table}
\end{center}

\subsection{Polarization cut sky experiment}\label{cut-sky experiment Simon}
The set-up of this second cut-sky experiment is the same as in the preceding section, except that we apply the Simons Observatory-motivated 37\% sky mask, see~\cite{Simon}, that we plot in Appendix~\ref{subsec:masks}, and set the noise rms to $\sigma= 0.28\mu K$ per pixel for both $EE$ and $BB$ spectra coefficients. Since we have very low SNR parameters we start binning the $BB$ multipoles at $\ell = 320$ into progressively wider bins. After binning is applied, we are left with $331$ bins out of the $512$ initial multipoles. For the non centered power spectrum sampling step, we assume a single bin for multipoles $2\leq \ell\leq 280$ followed by bins of width 1 for $280 < \ell \leq 331$.

As in the previous section, we make tuning runs of 10 parallel chains of 300 iterations. Then, we run all algorithms for 10 parallel chains of $10^{3}$ iterations, except for the Centered 1 and Centered overrelax cases, for which we run 10 parallel chains of length $10^{5}$. Again, following \cite{eriksen_power_2004}, we set the threshold for the PCG algorithm to $10^{-6}$, except for the ASIS RJPO algorithm, for which we set the threshold to $10^{-5}$.

\begin{figure*}[htpb]
\centering
\includegraphics[scale=0.5]{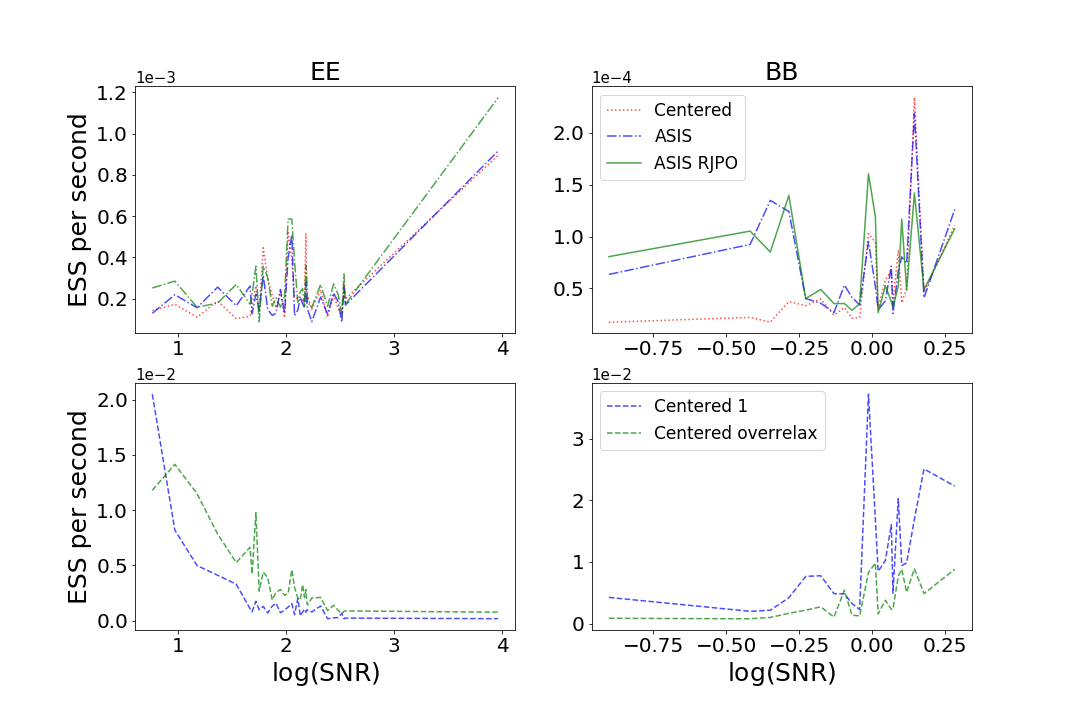}
\caption{Effective Sample Size per second against $\log(\SNR )$ for each algorithm and the cut-sky experiments. For the sake of clarity, we plot the results for the Centered 1 case, separately.}
\label{fig:EssSecSNRSimon}
\end{figure*}

Figure~\ref{fig:EssSecSNRSimon} shows the $\ESS$ per CPU second against $\log(\SNR )$. ASIS, ASIS RJPO and Centered algorithms tend to perform the same, except on the lower range of SNR where Centered algorithm is outperformed by ASIS and ASIS RJPO. We also note ASIS RJPO performs better than ASIS on the low SNR paremeters. 

Clearly, the Centered 1 and Centered overrelax variants outperform all other algorithms, sometimes by several orders of magnitude and over almost the entire range of SNR: that is because it is computationally very cheap compared to the other algorithms. Note however that its mixing properties degrade for extreme, either too high or too low, SNR. That is because the auxiliary step mixes worse on high SNR while the centered parametrization provides a bad mixing on low SNR.

Finally, Tables~\ref{TableCenteredESSSimonEE} and \ref{TableCenteredESSSimonBB} summarize the distribution of the ratios of ESS per second. On average, on $EE$, the Centered 1 algorithm performs 14 times better than the ASIS and the Centered ones, with a minimum of 0.17 and a maximum of 416. The few multipoles for which the $\ESS$ per second is worse than that of the ASIS and Centered cases are the ones corresponding to the highest SNR, where the auxiliary variable step mixes very badly. On $BB$, the Centered 1 variant performs on average 214 times better than the ASIS and Centered approaches, with a minimum at 5 and a maximum at 1147. Note that the Centered overrelax algorithm performs better than the Centered 1 one on $EE$ but not on $BB$ coefficients. However, it still performs much better on the $BB$ coefficients than the ASIS and Centered variants.

\begin{figure}[htpb]
    \centering
    \includegraphics[scale=0.24]{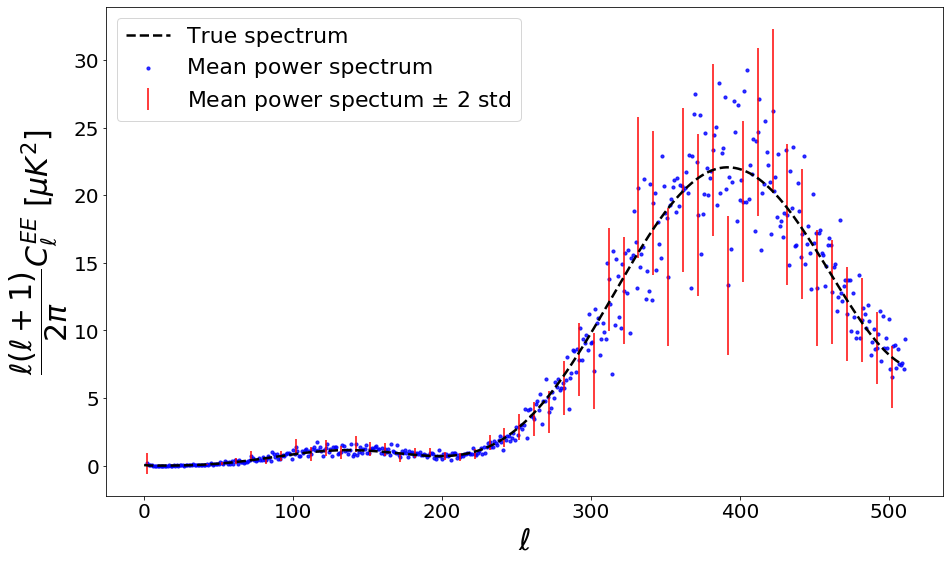}
    \includegraphics[scale=0.24]{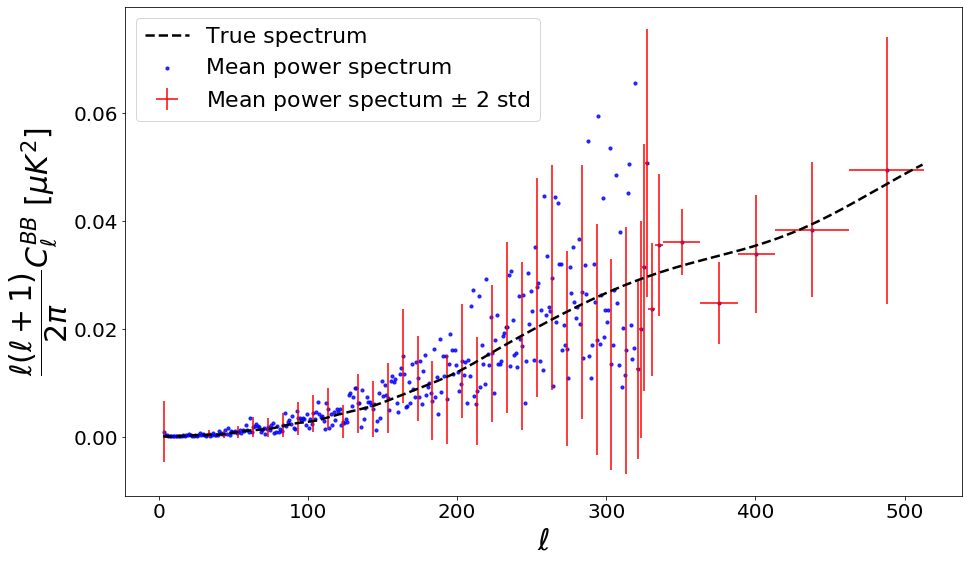}
    \caption{\textcolor{black}{An example of the constraints on the power spectra derived in the cut-sky experiment, using the Centered overrelax algorithm in the case of the ground-based experiment discussed in the text. The black dashed lines show the true input spectra. The blue points show the best power estimates in each bin equal to the mean power computed over the generated chains. The vertical bars of the crosses correspond to the mean plus/minus one standard deviation, and the horizontal bars the corresponding bins in $\ell$. We only plot the crosses for one every ten multipoles on the non binned part. We plot the crosses for every multipole on the binned part. The top panel shows the $EE$ power spectrum and the bottom -- the $BB$ one. The input CMB maps assumed the standard cosmological model with the assumed tensor-to-scalar ratio $r=0.001$.}}
    \label{fig:recoveredSpecCenteredOverrelaxSimon}
\end{figure}

Figure~\ref{fig:recoveredSpecCenteredOverrelaxSimon} shows the empirical mean posterior distribution of Centered overrelax with the two standard deviation interval. The solid black line denotes the true power spectrum.

\begin{center}
\begin{table}
\begin{tabular}{ |c|c|c|c|c|c|  } 
 \hline
 Algorithm &$5$th&$25$th&$50$th&$75$th&$95$th \\ [2ex] 
 \hline 
 ASIS & 0.549 & 0.738 & 0.991 & 1.248 & 1.703 \\ [1ex]\hline
 ASIS RJPO & 0.633 & 0.88 & 1.111 & 1.349 & 1.973\\ [1ex]\hline
 Centered 1 & 
1.172 & 2.819 & 5.469 & 10.991 & 60.119\\ [1ex]\hline
 Centered overrelax & 
3.854 & 7.75 & 12.709 & 23.646 & 93.642\\[1ex]\hline
\end{tabular}
\caption{\label{TableCenteredESSSimonEE} 
For each algorithm, percentiles of their $\ESS$ per second relative to the $\ESS$ per second of the Centered algorithm for $EE$ coefficients. These results are for the cut-sky experiment, see Section~\ref{cut-sky experiment Simon}.}
\end{table}
\end{center}

\begin{center}
\begin{table}
\begin{tabular}{ |c|c|c|c|c|c|  } 
 \hline
 Algorithm &$5$th&$25$th&$50$th&$75$th&$95$th \\ [2ex] 
 \hline 
 ASIS & 0.548 & 0.785 & 1.067 & 1.497 & 4.4 \\ [1ex]\hline
 ASIS RJPO & 0.621 & 0.982 & 1.239 & 1.819 & 4.922\\ [1ex]\hline
 Centered 1 & 75.717 & 129.257 & 183.94 & 279.193 & 516.07\\ [1ex]\hline
 Centered overrelax & 28.877 & 52.064 & 75.548 & 104.045 & 157.7\\[1ex]\hline
\end{tabular}
\caption{\label{TableCenteredESSSimonBB} 
For each algorithm, percentiles of their $\ESS$ per second relative to the $\ESS$ per second of the Centered algorithm for $BB$ coefficients. The shown results are for the cut-sky experiment, see Section~\ref{cut-sky experiment Simon}.}
\end{table}
\end{center}

\section{Conclusion}\label{Conclusion}

We have discussed and compared a number of the Gibbs samplers implemented in the context of the CMB power spectrum estimation.

Two of the studied cases, the centered Gibbs~\cite{eriksen_joint_2008} and non centered Gibbs~\cite{jewell_application_2004} samplers, have been previously applied to the inference of the power spectrum of the CMB signal. While both the variants have been demonstrated to be feasible, they have been also found to be computationally very demanding. Two main reasons behind it have been identified~\cite{eriksen_joint_2008, jewell_application_2004}. First, both these algorithms display poor sampling efficiency, with  
the centered Gibbs failing on the low SNR coefficients and the non-centered Gibbs on the high SNR coefficients. Second, both these algorithms require significant computations for every sky signal sample due to the need for solving the constrained realization system of equations. We have elaborated on both these factors from the theoretical perspective and demonstrated them via numerical experiments.

We have subsequently proposed a number of possible extensions aiming at improving the overall performance of these two methods.

First, we have looked at improving the sampling efficiency of the standard algorithms. To this end, we have introduced the interweaving concept proposed earlier in the statistical literature, and implemented it in the CMB power spectrum estimation context to improve on the mixing of the over the entire range of signal-to-noise ratio, enabling a more efficient sampling of the entire power spectrum. While potentially promising the improvement comes at the cost of increased computational time per sample.

Second, we have looked at the ways of lowering the cost of single sample computations via statistical means. We have considered two approaches here. 
Our first proposal, the RJPO algorithm allows for approximate solutions to the constrained realization problem without introducing biases to the final results and does not increase the sample autocorrelation length, if proper tuning is ensured.
Our second proposal alleviates the need for solving the constrained realization system altogether by introducing an auxiliary variable. The algorithm is easy to implement and tuning free, but comes at the price of increased dimensionality of the problem.

We have compared and studied all these variants on simulated CMB maps with full-sky, nearly full-sky, and limited cut-sky coverage.

{\color{black}We have found that the Centered overrelax and the Centered 1 variants are consistently performing the best out of the all algorithms considered here and for all studied simulated cases.
The performance gain is often very significant. In particular, } in the cases with full and nearly full sky coverage the Centered overrelax algorithm has performed, on average, an order of magnitude better on $EE$ and two orders of magnitude better on $BB$ in terms of $\ESS$ per second than the other algorithms.

This Centered overrelax algorithm exhibits, however, some drawbacks: for very low or very high signal-to-noise ratio, coefficients it produces long autocorrelations. For the very low ratios, it is because of the centered parametrization, while for the very high ones, it is because of the bad mixing of the augemented Gibbs sampler step. To solve the first problem, we face a trade-off: we may improve on the mixing by using a non-centered step on the lowest SNR ratio coefficients, but this would increase the cost of the algorithm. To solve the second problem, we would need to find a better mixing algorithm for the constrained realization step, or to find more efficient ways to solve the system.

\textcolor{black}{Note that Centered 1 tends to have the same shortcomings, except that the bad mixing for the very high SNR is even worse, since we are not using the overrelexation strategy on top of the auxiliary scheme. For the very low SNR, centered 1 also struggle because of its centered parametrization. It mixes somewhat better than centered overrelax because it is a bit cheaper.}

We also note that another MCMC algorithm addressing these problems and seemingly efficient for the entire range of SNR ratios has been developed in~\cite{racine_cosmological_2016}. However, to our knowledge, this algorithm has only been used to make inference on the cosmological parameters instead of the power spectrum. In addition, it requires two resolutions of the high dimensional constrained realization system and is thus very costly and requires the tuning of proposal distributions. 

{\color{black} The experiments used in this work were applied to CMB-only sky maps with rather simply noise. In practice, the CMB power spectra will have to be derived from multi-frequency data sets where the sky signal comprises cosmological, astrophysical, such as foregrounds, and often environmental effects, e.g., atmosphere, as well as instrumental contributions. The sampling algorithms proposed here will have to be extended to account for such effects, what will require adapting them to more complex and involved likelihoods, e.g.,~\cite{Katayama2011, Switzer2016, BeyondPlanck2022, Stompor2009}. A significant progress in this direction in the context of the standard, centered Gibbs algorithm has been demonstrated in~\cite{commander2022,commander2022a}, which however continue being affected by significant numerical cost. The algorithms proposed and studied in this work should be directly relevant in the context of overcoming the current limitations of these implementations.  We leave this for future work.}
\\
\section{Acknowledgements}
We thank Clément Leloup for comments, careful reading of the manuscript, and stimulating discussions. This work  was performed in the context of the B3DCMB project and benefited from its inspiring, multidisciplinary environment. We acknowledge the use of \texttt{healpy}~\cite{Zonca2019}, \texttt{Healpix}~\cite{Gorski2005}, \texttt{numpy}~\cite{harris2020array} and \texttt{maplotlib}f~\cite{Hunter:2007}.
This research was supported by the French National Research Agency (ANR) grant, ANR-B3DCMB, (ANR-17- CE23-0002). JE and RS acknowledge additional support of the ANR-BxB grant (ANR-17-CE31-0022).

\bibliography{myBibDefinitive, introductionThesis, CubeThinning} 
\bibliographystyle{ieeetr}

\appendix

\section{Improper priors}\label{PriorsAppendix}
In this appendix we show that in the case of a full sky observation and a noise matrix proportional to identity $\NoiseMatrix = \alpha \boldsymbol{I}$, using a flat prior over the power spectrum lead to a proper posterior distribution while Jeffrey's prior leads to an improper posterior distribution.

Since we assume full sky coverage and a noise matrix proportional to identity, we can rewrite Model \ref{CenteredModel} in harmonic domain:
\[
d = s + n 
\]
where $s$ is the signal map expressed in the spherical harmonics basis - the vector of $(a_{l,m})_{2\leq l \leq \lmax, 0 \leq m \leq l}$ coefficients, that is $s \sim \mathcal{N}(0, \C(\{C_{\ell}\}))$. We also now have $n \sim \mathcal{N}(0, \Beam^{-2}\alpha w)$ where $w= \dfrac{4\pi}{\Npix}$, $\alpha$ being the noise matrix in spherical harmonics basis. It follows that $d$ is the observed skymap expressed in harmonic domain too.

In this case, the likelihood straightforwardly writes as:

\begin{multline}
    \mathcal{L}(d|\{C_{\ell}\}) = \prod_{\ell = 2}^{\lmax} \dfrac{\exp\left(-(1/2) \dfrac{||d_{\ell}||_{2}^{2}}{C_{\ell} + b_{\ell}^{-2}\alpha w}\right)}{|C_{\ell} + b_{\ell}^{-2}\alpha w|^{(2\ell+1)/2}}\\\times \mathbf{1}_{\{C_{\ell} > 0\}}
\end{multline}

Let's suppose we are using a flat prior on the power spectrum. In this case, we have $\pi(\{C_{l}\}|d) \propto \mathcal{L}(d | \{C_{l}\})$. Then, doing the following change of variable: $y_{l} = C_{l} + b_{l}^{-2}\alpha w$ we have:
\begin{multline*}
\int_{0}^{\infty} \mathcal{L}(d | \{C_{l}\}) dC_{2} \dots dC_{\lmax} \propto \int_{0}^{\infty} \prod_{l = 2}^{\lmax} p_{\gamma}(y_{l}; \alpha_{l}, \beta_{l})\\\times \mathbf{1}_{\{y_{l} > b_{l}^{-2}\alpha w\}} dy_{2} \dots dy_{\lmax}
\end{multline*}
up to a positive multiplicative constant.
Here $p_{\gamma}$ means inverse Gamma distribution with parameters $\beta_{l} = \dfrac{||d_{l}||_{2}^{2}}{2}$ and $\alpha_{l} = \dfrac{2l-1}{2}$. Since $\mathbf{1}_{\{y_{l} > b_{l}^{-2}\alpha w\}} \leq \mathbf{1}_{\{y_{l} > 0\}}$, we have:
\begin{multline*}
\int_{0}^{\infty} \mathcal{L}(d | \{C_{l}\}) dC_{2} \dots dC_{\lmax} \lesssim \int_{0}^{\infty} \prod_{l = 2}^{\lmax} p_{\gamma}(y_{l}; \alpha_{l}, \beta_{l})\\ dy_{2}\times \dots \times dy_{\lmax}
\end{multline*}
And the right-hand term of this equation is integrable as the product of independant inverse Gamma densities. Hence, the posterior distribution we obtain with a flat prior is proper.

Now, with Jeffrey's prior $p(\{C_{l}\}) = \prod_{l=2}^{\lmax} \dfrac{1}{C_{l}}$, things are different:
\begin{multline*}
\int_{0}^{1} \mathcal{L}(d | \{C_{l}\})p(\{C_{l}\}) dC_{2} \dots dC_{\lmax} = \int_{0}^{1} \prod_{l = 2}^{\lmax}\\\times \dfrac{\exp\left(-(1/2) \dfrac{||d_{l}||_{2}^{2}}{C_{l} + b_{l}^{-2}\alpha w}\right)}{|C_{l} + b_{l}^{-2}\alpha w|^{(2l+1)/2}}\dfrac{1}{C_{l}} \mathbf{1}_{\{C_{l} > 0\}}.
\end{multline*}
But, on $]0, 1]$ and for any $l\in \{2, \dots, \lmax\}$ we have:
\[
\exp\left(-(1/2) \dfrac{||d_{l}||_{2}^{2}}{C_{l} + b_{l}^{-2}\alpha w}\right) \geq \exp\left(-(1/2) \dfrac{||d_{l}||_{2}^{2}}{b_{l}^{-2}\alpha w}\right)
\]
and
\[
\dfrac{1}{|C_{l} + b_{l}^{-2}\alpha w|^{(2l+1)/2}} \geq \dfrac{1}{|1 + b_{l}^{-2}\alpha w|^{(2l+1)/2}}
\]
Hence we have

\begin{multline*}
\int_{0}^{1} \mathcal{L}(d | \{C_{l}\})p(\{C_{l}\}) dC_{2} \dots dC_{\lmax} \gtrsim \int_{0}^{1} \prod_{l=2}^{\lmax} \dfrac{1}{C_{l}} dC_{2}\\\times
\dots \times dC_{\lmax}
\end{multline*}
And obviously the right-hand side diverges to infinity. Since the integrand of the left-hand side is positive on $]0, \infty[$, this proves that the posterior distribution is improper if we use Jeffrey's prior on the power spectrum.

\section{Signal-to-noise ratio}\label{appendix:snr}
\begin{figure}[htpb]
    \centering
    \includegraphics[scale=0.19]{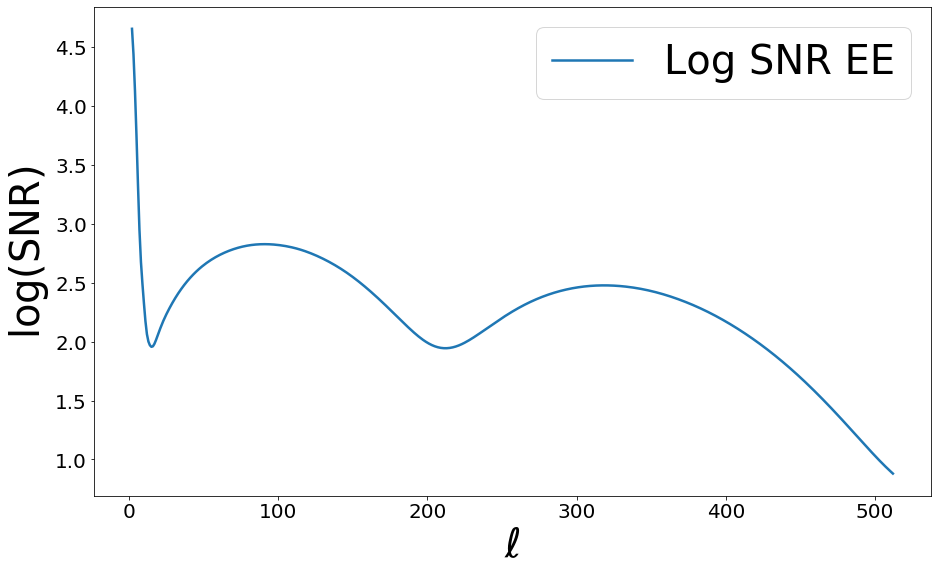}
    \includegraphics[scale=0.19]{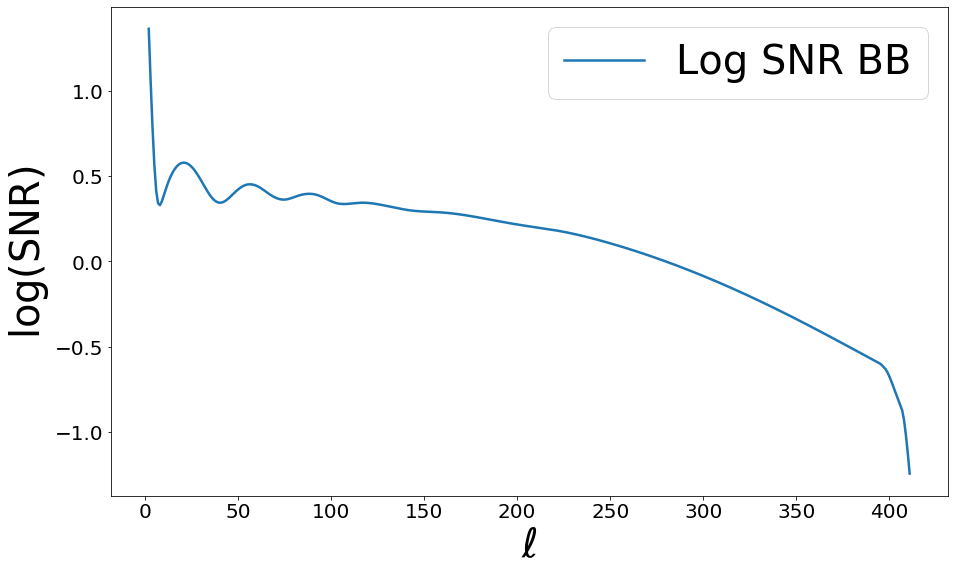}
    \caption{Logarithm of the signal-to-noise ratio for the full sky experiment of Section~\ref{fullskyExp}. Top: $EE$ spectrum coefficients. Bottom: $BB$ ones.}
    \label{fig:fullSkySNR}
\end{figure}

\textcolor{black}{The signal to noise ratio is defined according to equation (\ref{eq:snr}). When we observe the entire sky and when the noise matrix is proportional to the identity matrix: $\NoiseMatrix = \alpha^2 I$, the noise power spectrum is given by:
\[
N_{\ell} = \alpha^{2}\dfrac{4\pi}{\Npix b^{2}_{\ell}}
\]
where the beam function $b_{\ell}$ is given by:
\[
b_{\ell} = \exp\{-\ell(\ell+1)\sigma^{2}_{\mathrm{FWHM}}/(8\log(2)\}
\]
with $\sigma_{\mathrm{FWHM}}\in\R$. Because of the exponential drop of the beam function, the noise power spectrum sharply increases with $\ell$.\\
We plot the SNR as a function of $\ell$ for the full sky experiment described in Section~\ref{fullskyExp}, see Figure~\ref{fig:fullSkySNR}.}

\section{Mixing}\label{mixing}
\begin{figure}[htpb]
\centering
\includegraphics[scale=0.35]{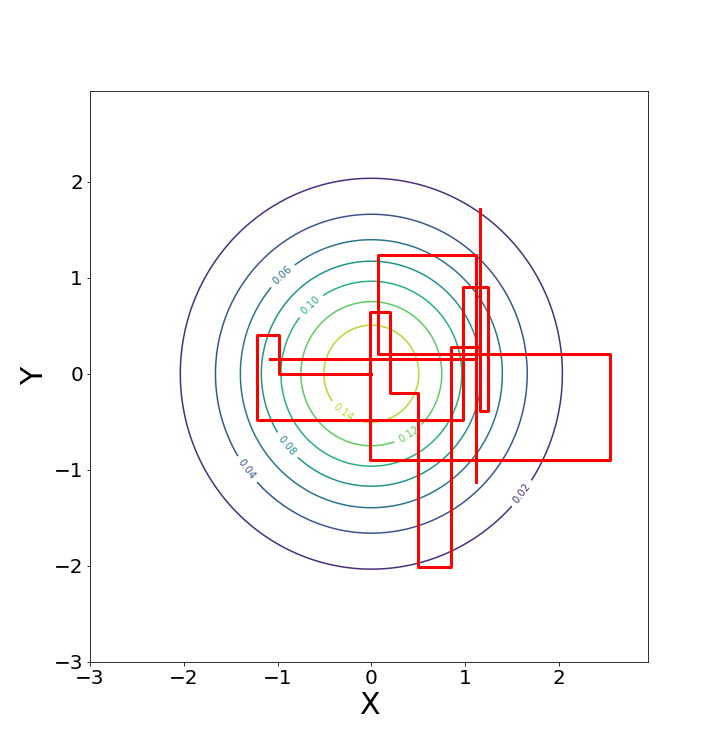}
\caption{Trace plot, in red, of the Gibbs sampler targeting the joint distribution of $(X,Y)$ for $\rho=0$, described in Appendix~\ref{mixing}. The circles are the level sets of the normal distribution.}
\label{fig:rho0tracePlots}
\end{figure}

\textcolor{black}{In this appendix we provide a intuitive understanding of what we call the "mixing" of a MCMC algorithm.\\
As a toy example, suppose we wish to sample the Gaussian vector $(X, Y)$ with mean zero and covariance matrix:
\begin{equation}
\boldsymbol{\Sigma}= 
\begin{pmatrix} 
1 & \rho \\
\rho & 1\\
\end{pmatrix}
\end{equation} 
where $\rho\in ]-1, 1[$.\\
Now we set $\rho = 0$ and we use a Gibbs sampler to sample from this distribution. We can plot the trajectory of the algorithm, see Figure~\ref{fig:rho0tracePlots}.}

\begin{figure}[htpb]
\centering
\includegraphics[scale=0.35]{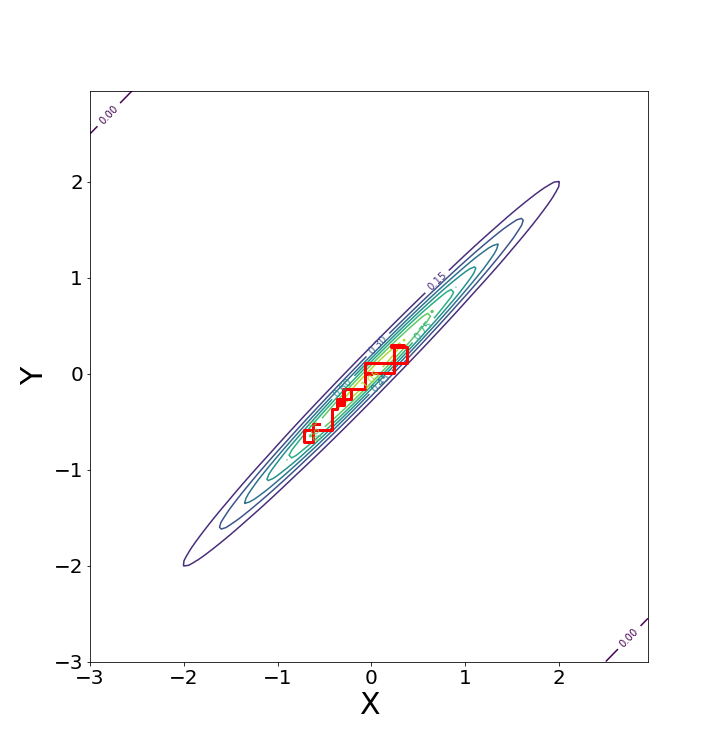}
\caption{Trace plot, in red, of the Gibbs sampler targeting the joint distribution of $(X,Y)$ for $\rho=0.99$, described in Appendix~\ref{mixing}. The circles are the level sets of the normal distribution.}
\label{fig:rho099tracePlots}
\end{figure}

\textcolor{black}{We can also suppose that $\rho=0.99$, in which case we get another trace plot, see Figure~\ref{fig:rho099tracePlots}.}

\begin{figure}[htpb]
    \centering
    \includegraphics[scale=0.35]{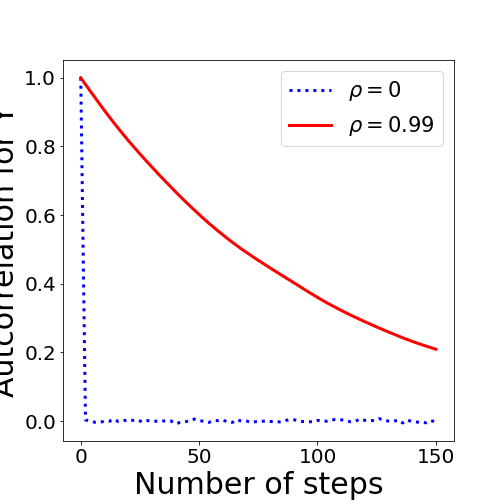}
    \caption{Autocorrelation plots of the Gibbs sampler, described in Appendix~\ref{mixing}, for $Y$ and for $\rho=0$ and $\rho=0.99$.}
    \label{fig:autocorY}
\end{figure}

\begin{figure}[htpb]
    \centering
    \includegraphics[scale=0.35]{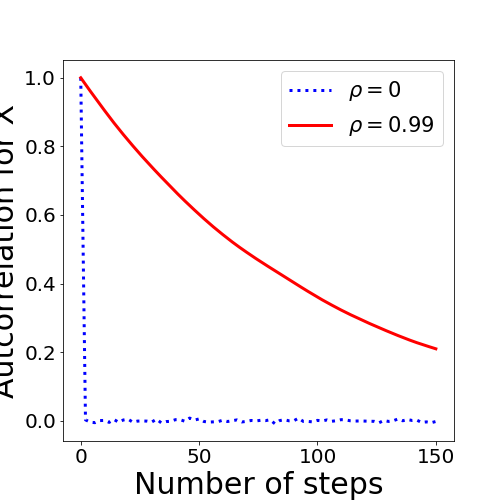}
    \caption{Autocorrelation plots of the Gibbs sampler, described in Appendix~\ref{mixing}, for $X$ and for $\rho=0$ and $\rho=0.99$.}
    \label{fig:autocorX}
\end{figure}

\textcolor{black}{We plot the autocorrelations of the Gibbs sampler for $X$ and $Y$ in Figure~\ref{fig:autocorX} and Figure~\ref{fig:autocorY}.}

\textcolor{black}{We can see on Figures~\ref{fig:rho0tracePlots} and \ref{fig:rho099tracePlots} that the Gibbs sampler with $\rho=0$ explores the target distribution much more efficiently than when $\rho=0.99$. We can also see on Figure~\ref{fig:autocorX} and Figure~\ref{fig:autocorY} that the autocorrelations are much longer when $\rho =0.99$ than when $\rho = 0$. More precisely, the Gibbs sampler samples independently when $\rho = 0$ while when $\rho = 0.99$, the sampled points are still correlated after 150 steps. When an algorithm explores the target distribution and shows low autocorrelations like the Gibbs sampler when $\rho = 0$, we say it is mixing well. On the contrary, when an algorithm behaves like the Gibbs sampler when $\rho = 0.99$, we say it is mixing badly. Here, the term "mixing" does not have a precise definition and we use it loosely.}\\

\textcolor{black}{Even though we use the term "mixing" loosely, we can still characterize the convergence of a Markov chain with state space $\mathcal{X}$, invariant distribution $\pi$ and transition kernel $P$. We usually want our Markov chain to converge geometrically to the invariant distribution $\pi$, that is:
\begin{equation}\label{def:geometicRate}
    ||P^{n}(x, \mathrm{d}y) - \pi(\mathrm{d}y)||_{\mathrm{TV}} \leq C r^{n}
\end{equation}
for any $x\in\mathcal{X}$, where $C >0$ and $r\in[0, 1)$ are constants. The constant $r$ is called the geometric rate of convergence.
}

\section{Experiments}
\subsection{Full-sky polarization experiment}
\label{full-sky appendix}
\begin{figure*}[htpb]
\centering
\includegraphics[scale=0.35]{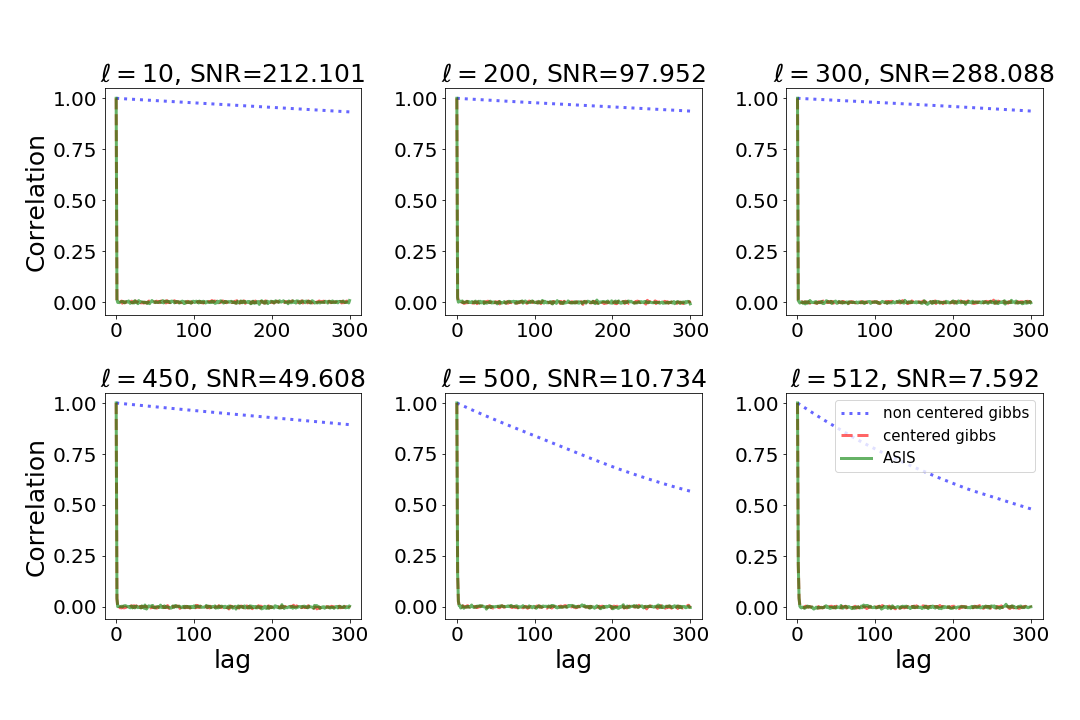}
\caption{Examples of autocorrelations for the EE spectrum coefficients for the full sky experiment, Section~\ref{fullskyExp}.}
\label{fig:autocorr_EE}
\end{figure*}

\begin{figure*}[htpb]
\centering
\includegraphics[scale=0.35]{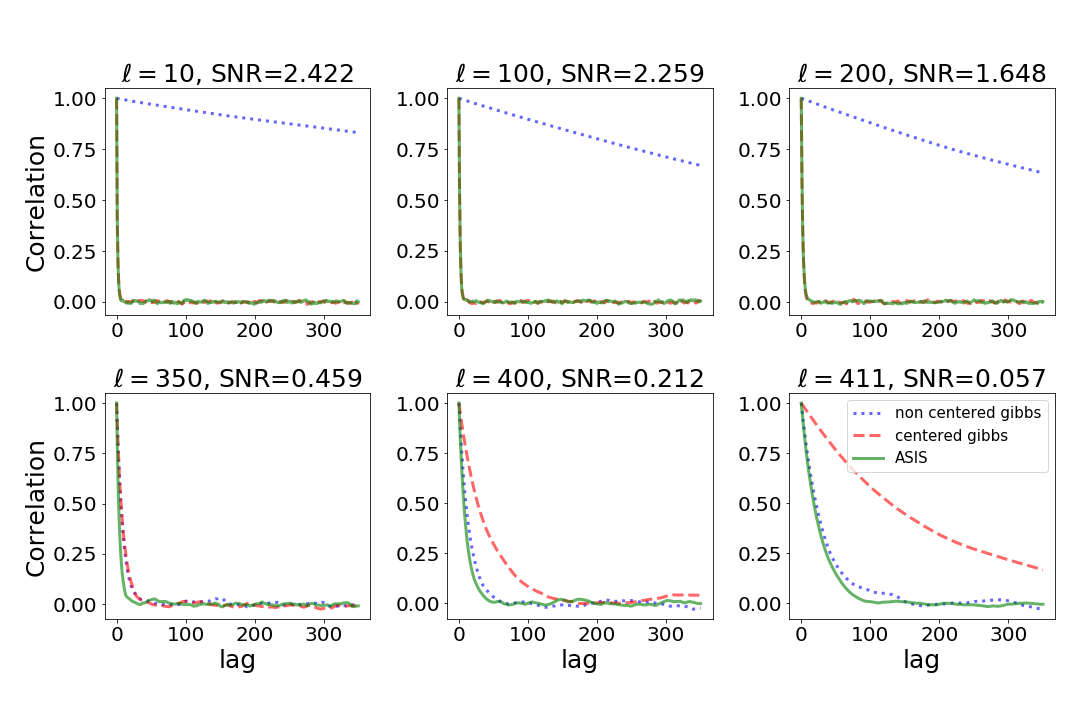}
\caption{Examples of autocorrelations for the BB spectrum coefficients for the full sky experiment, Section~\ref{fullskyExp}.}
\label{fig:autocorr_BB}
\end{figure*}

\begin{figure*}[htpb]
\centering
\advance\leftskip-0.8cm
\includegraphics[scale=0.4]{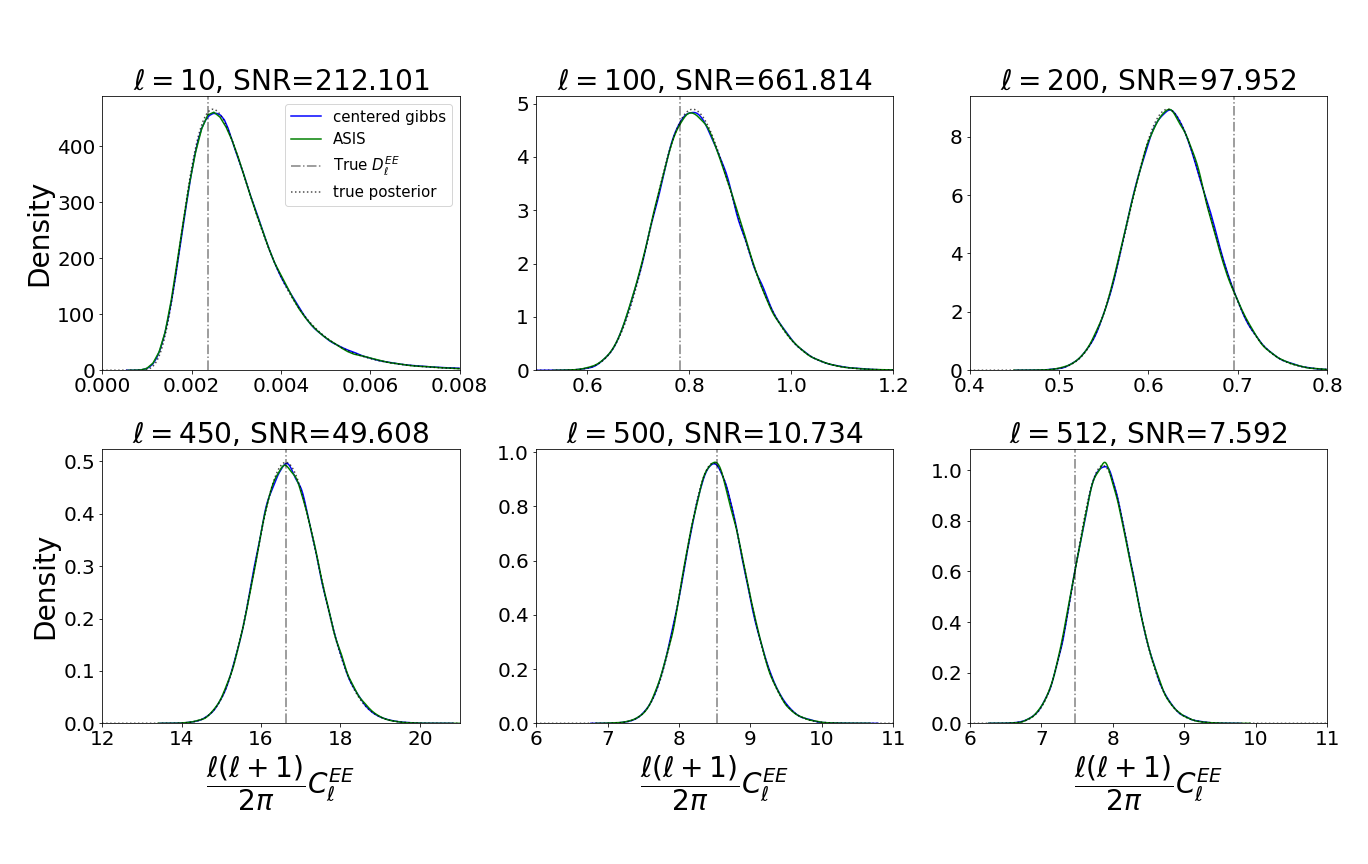}
\caption{Examples of kernel density estimation of histograms for EE spectrum coefficients for the full sky experiment, Section~\ref{fullskyExp}. For readability and since the mixing of the non centered Gibbs is bad, we do not include its histograms on this figures. }
\label{fig:histo_EE}
\end{figure*}

\begin{figure*}[htpb]
\centering
\advance\leftskip-0.8cm
\includegraphics[scale=0.4]{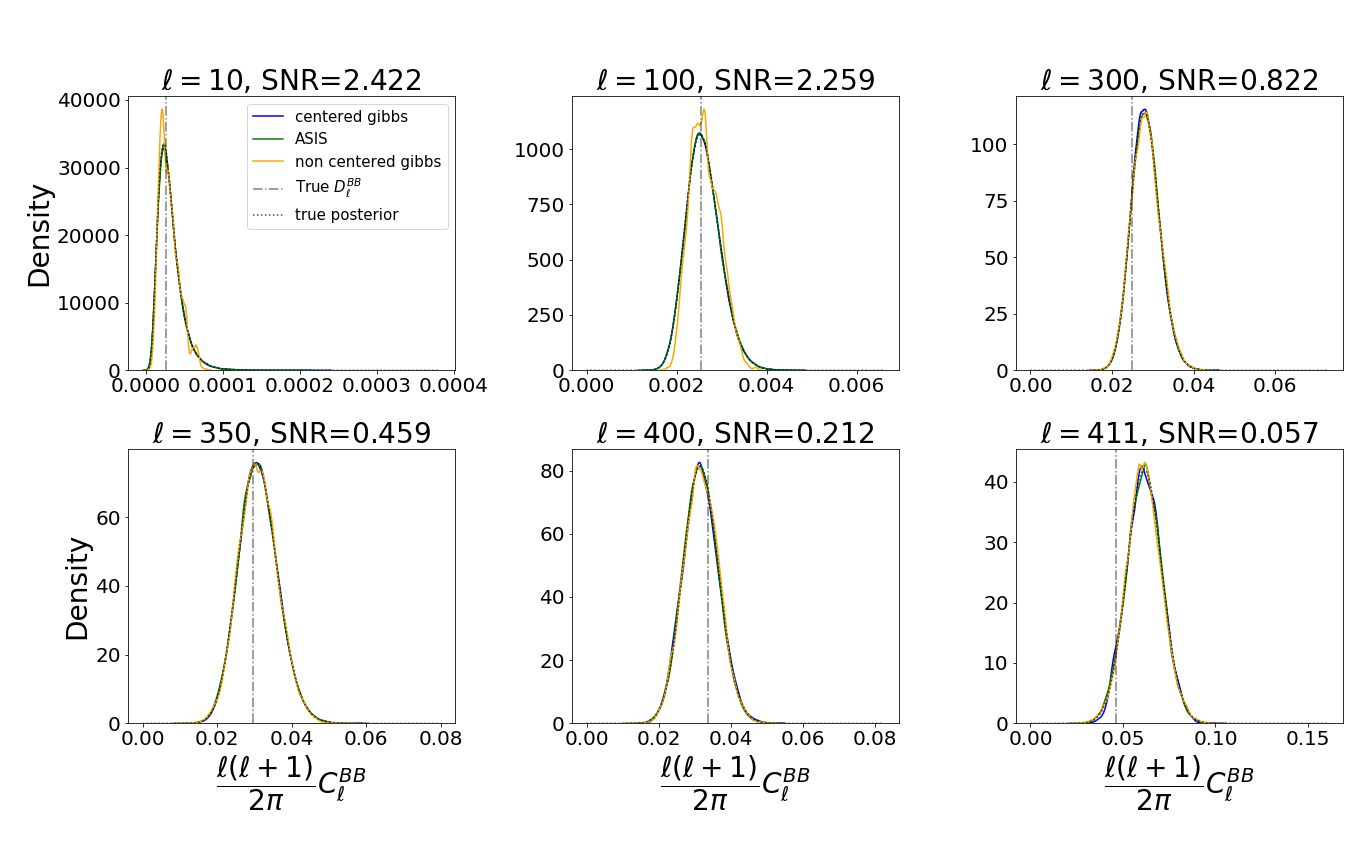}
\caption{Examples of kernel density estimation of histograms for the BB coefficients for the full sky experiment, Section~\ref{fullskyExp}. For readability and since the mixing of the non centered Gibbs is bad, we do not include its histograms on this figures.}
\label{fig:histo_BB}
\end{figure*}

In this appendix we show histograms and autocorrelation plots that we obtained running the full-sky experiment described \ref{fullskyExp} on Figures~\ref{fig:autocorr_EE} to \ref{fig:histo_BB}. All these figures confirm our analysis of Section~\ref{interweaving} and the results of Section~\ref{fullskyExp}: the interweaving algorithm performs as good as the centered Gibbs on high SNR coefficients and as good as the non-centered Gibbs on low SNR ones. Note also that the kernel density estimation of the histograms of interweaving matches almost perfectly the true posterior marginals for any signal-to-noise ratio.

\subsection{Sky masks}\label{subsec:masks}
In this appendix we provide plots of the two sky maps used in the experiment section.
\begin{figure*}[htpb]
\centering
\includegraphics[scale=0.5]{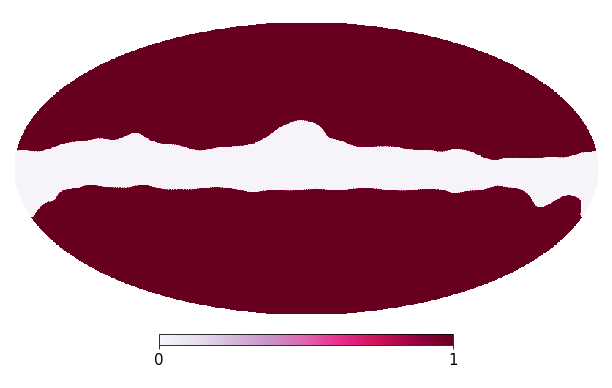}
\caption{Planck sky mask used for the nearly full-sky experiment described in Section~\ref{cutSkyPlanckMask}. This mask covers roughly 80\% of the sky.}
\label{fig:planckMask}
\end{figure*}

\begin{figure*}[htpb]
\centering
\includegraphics[scale=0.5]{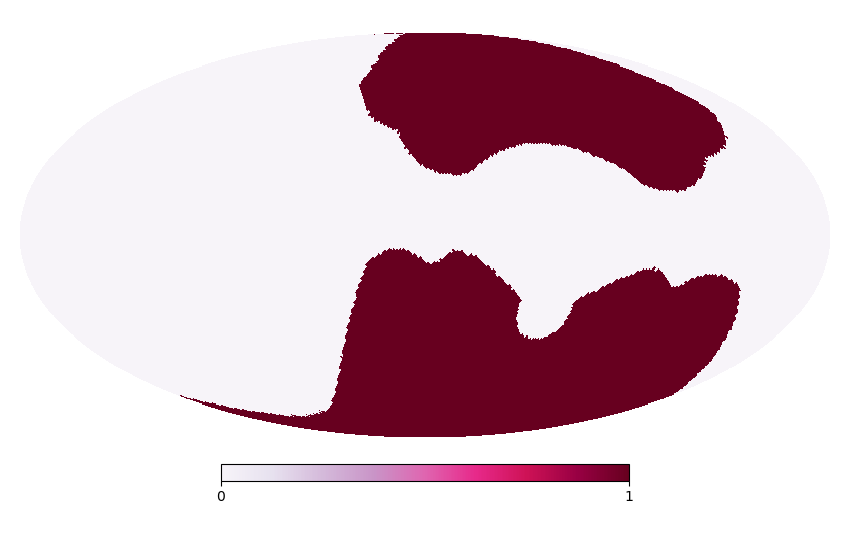}
\caption{A sky mask used for the second cut sky experiment described in Section~\ref{cut-sky experiment Simon} and motivated by the Simons Observatory like experiment. This mask covers roughly 35\% of the sky.}
\label{fig:simonMask}
\end{figure*}

\subsection{A first cut-sky polarization experiment}
\label{cut-sky Planck appendix}
\begin{figure*}[htpb]
\centering
\includegraphics[scale=0.4]{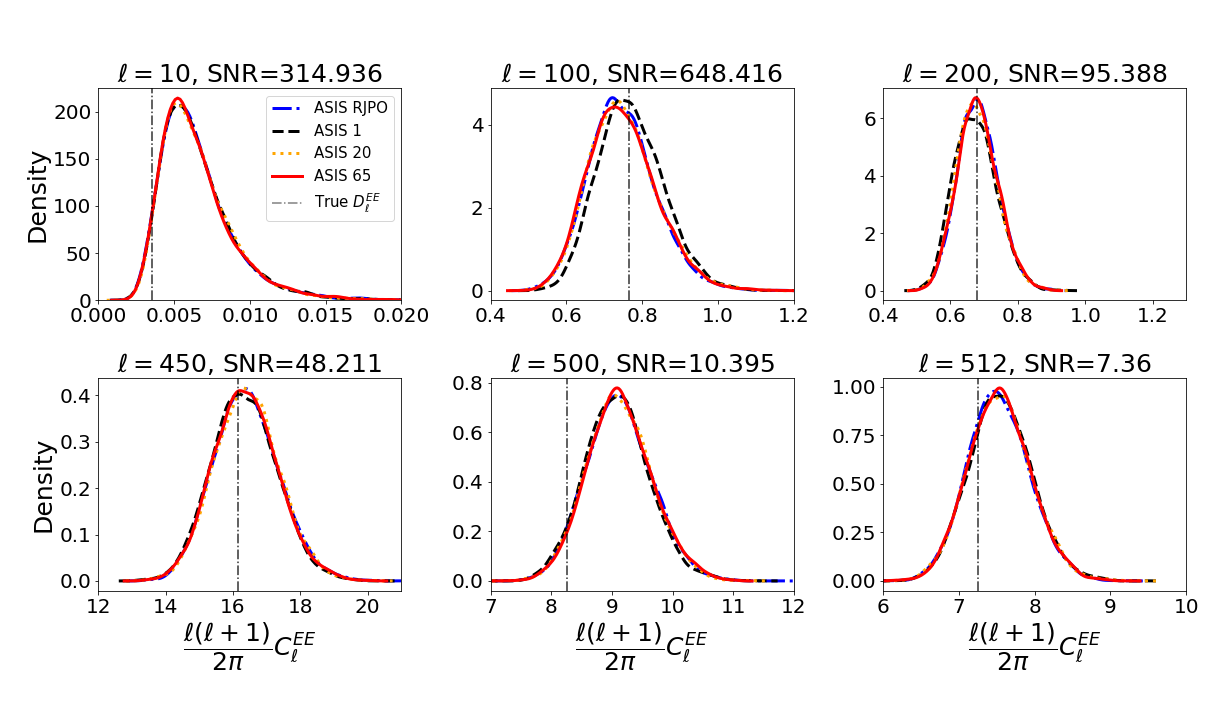}
\caption{Kernel density estimation of marginals for a sample of multipoles for the EE coefficients for the cut-sky experiment, Section~\ref{cutSkyPlanckMask}.}
\label{fig:kdePlotASISEE}
\end{figure*}

\begin{figure*}[htpb]
\centering
\includegraphics[scale=0.4]{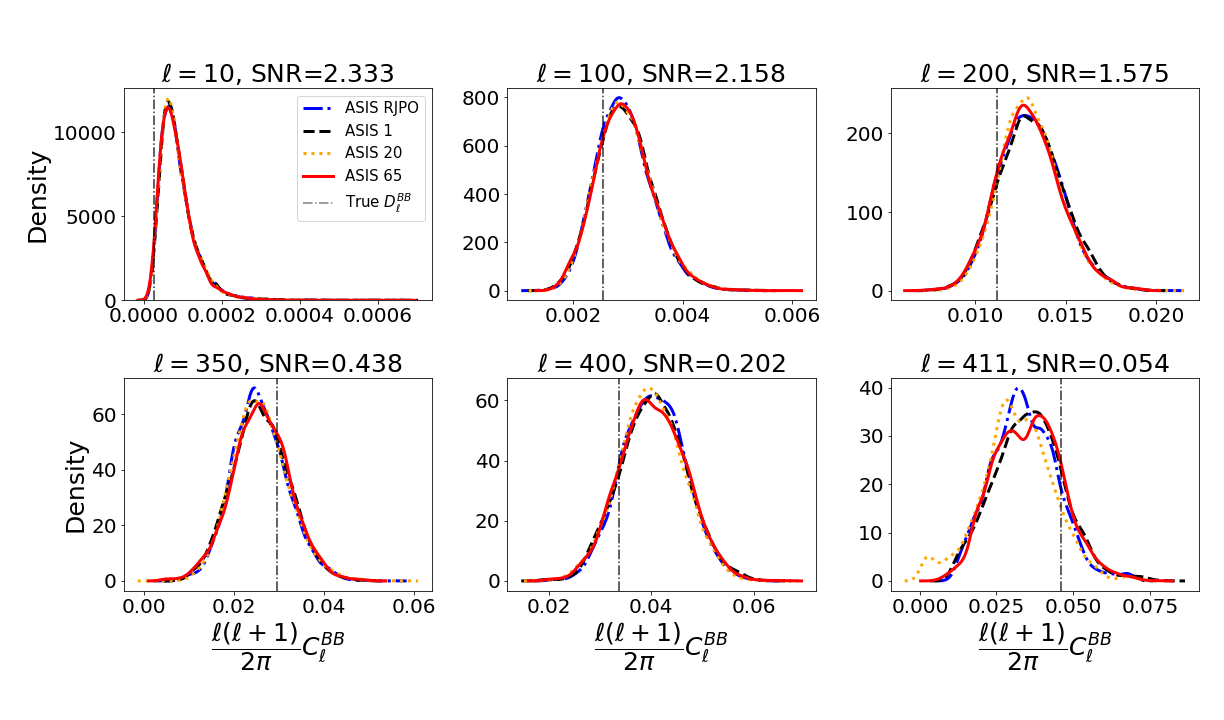}
\caption{Kernel density estimation of marginals for a sample of multipoles for the BB spectrum for the cut-sky experiment, Section~\ref{cutSkyPlanckMask}.}
\label{fig:kdePlotASISBB}
\end{figure*}

\begin{figure*}[htpb]
\centering
\includegraphics[scale=0.3]{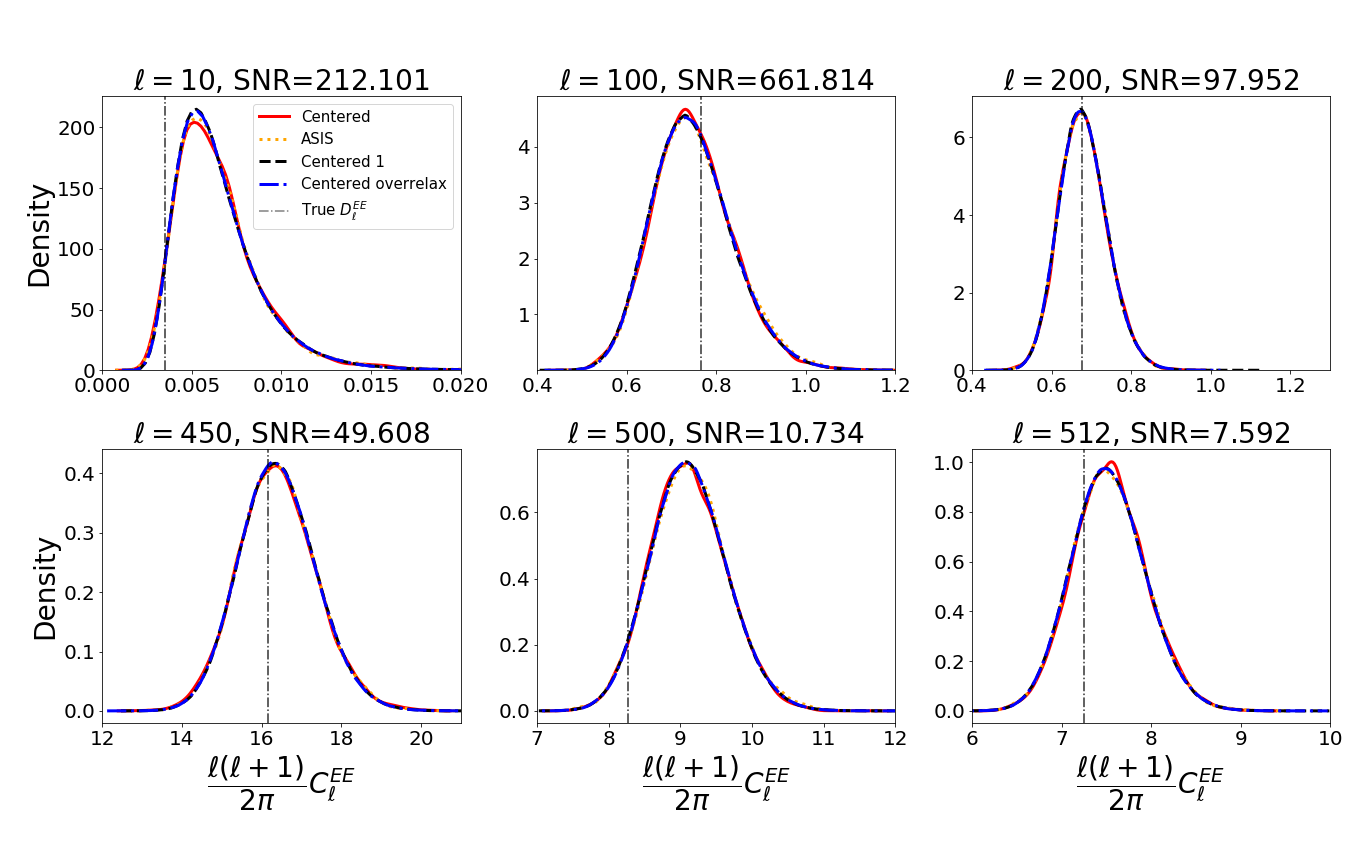}
\caption{Kernel density estimation of marginals for a sample of multipoles for the EE spectrum coefficients for the cut-sky experiment, Section~\ref{cutSkyPlanckMask}.}
\label{fig:kdePlotCenteredEE}
\end{figure*}

\begin{figure*}[htpb]
\centering
\includegraphics[scale=0.3]{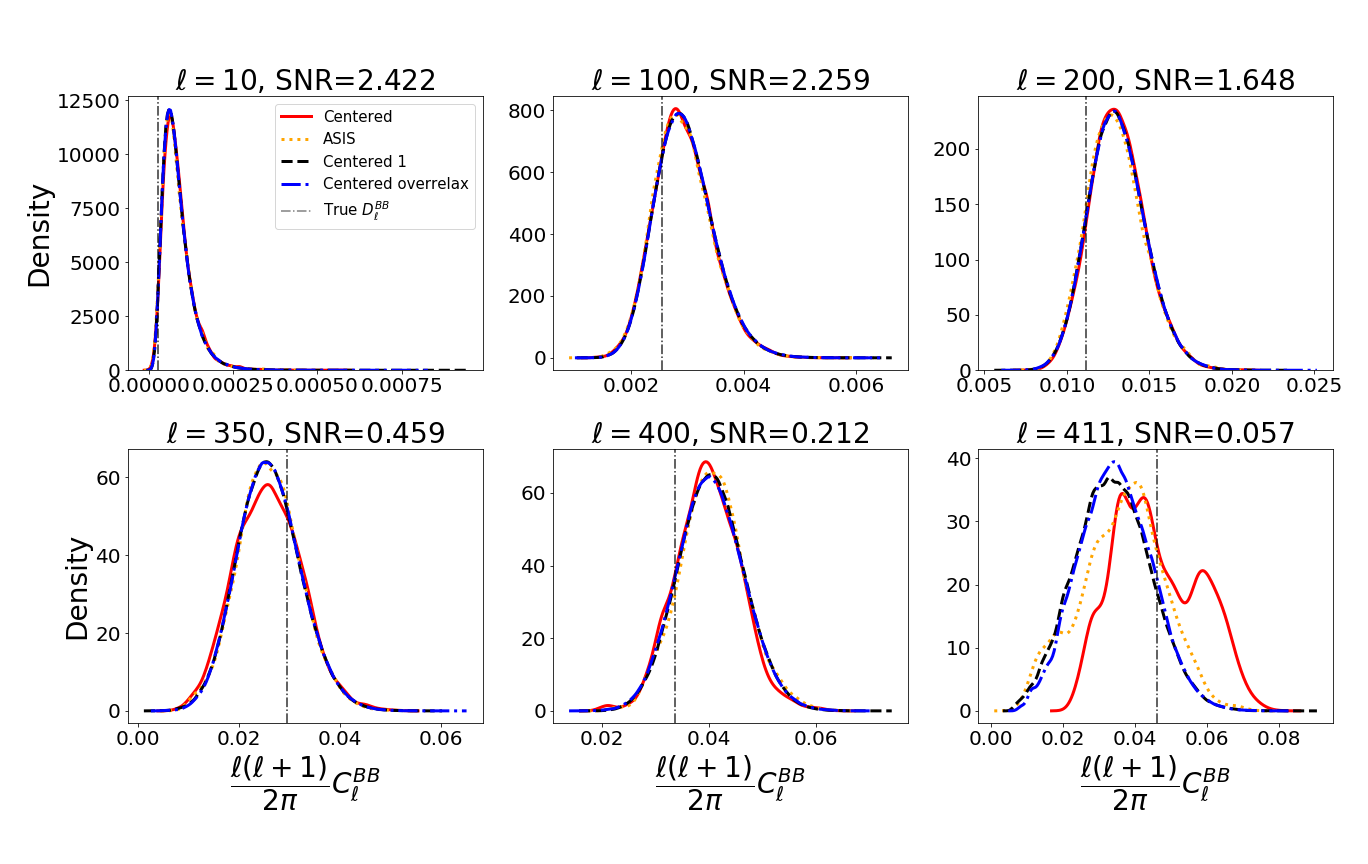}
\caption{Kernel density estimation of marginals for a sample of multipoles of the EE spectrum coefficients for the cut-sky experiment, Section~\ref{cutSkyPlanckMask}.}
\label{fig:kdePlotCenteredBB}
\end{figure*}

This appendix provides kernel density estimation based on the histograms of the histograms used in Section~\ref{cutSkyPlanckMask}. See Figures~\ref{fig:kdePlotASISEE} to \ref{fig:kdePlotCenteredBB}.

\subsection{A second cut-sky polarization experiment}
\label{cut-sky Simon appendix}
\begin{figure*}[htpb]
\includegraphics[scale=0.3]{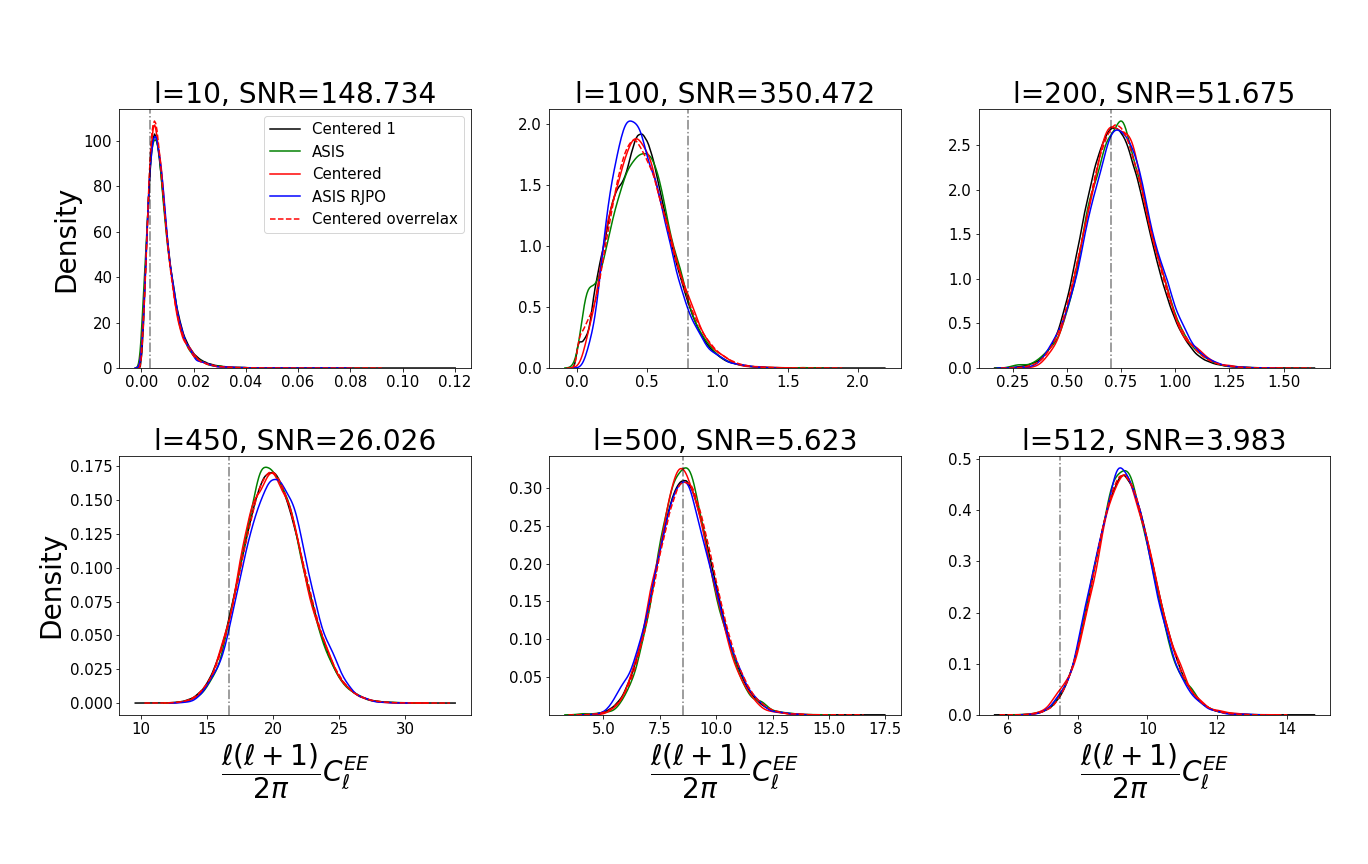}
\caption{Kernel density estimation of the posterior density of the $EE$ spectrum coefficients for the  cut-sky experiment, Section~\ref{cut-sky experiment Simon}.}
\label{fig:kdePlotEESimon}
\end{figure*}

\begin{figure*}[htpb]
\centering
\includegraphics[scale=0.3]{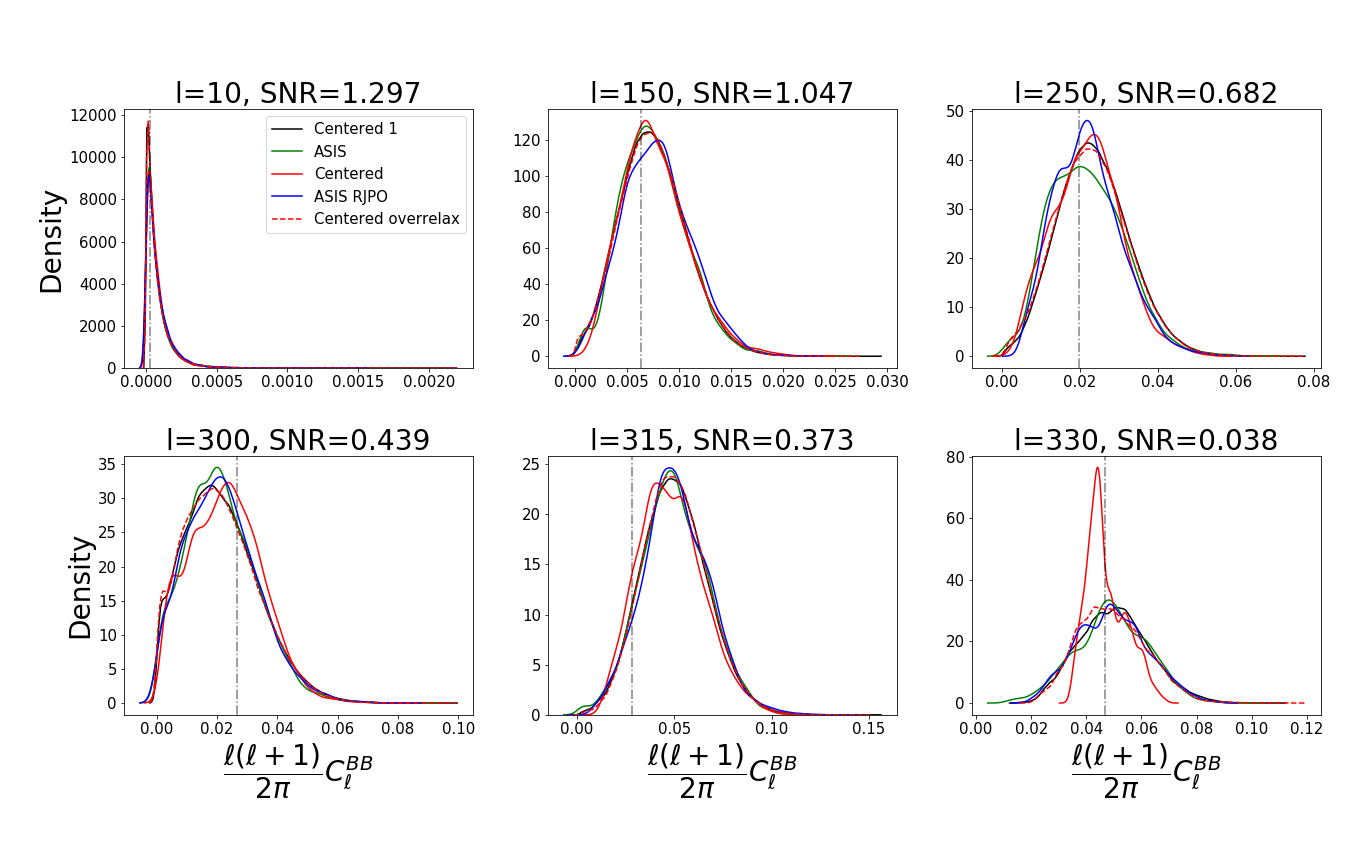}
\caption{Kernel density estimation of the posterior density of the $BB$ spectrum for the cut-sky experiment, Section~\ref{cut-sky experiment Simon}.}
\label{fig:kdePlotBBSimon}
\end{figure*}

This appendix provides kernel density estimation based on the histograms obtained in Section~\ref{cut-sky experiment Simon}. See Figures~\ref{fig:kdePlotEESimon} and \ref{fig:kdePlotBBSimon}.

Note that for the lowest SNR $BB$ coefficients, Centered gives an irrelevant estimate of the posterior density while Centered 1 gives a result in agreement with ASIS and ASIS RJPO: even though Centered 1 suffers from the centered parametrization, thanks to its low computational cost, we are able to perform enough iterations to have a reliable estimate. Which is not the case of Centered because of its high computational cost.  

\end{document}